\documentclass[aip,jcp,amsmath,amssymb,preprint]{revtex4-1}

\usepackage{graphicx}% Include figure files
\usepackage{dcolumn}% Align table columns on decimal point
\usepackage{multirow}
\usepackage{bm}
\usepackage{natbib}
\usepackage{times}
\usepackage{mathptmx}
\usepackage[version=3]{mhchem}  % this is a great package for formatting chemical reactions
\usepackage{url}
\usepackage{braket}
\usepackage{tabularx}
\usepackage{rotating}
\usepackage{multirow}
\usepackage{booktabs}

\newcolumntype{Y}{>{\centering\arraybackslash}X}

\begin{document}

\title{Real space electrostatics for multipoles. III. Dielectric Properties}

\author{Madan Lamichhane}
\affiliation{Department of Physics, University
of Notre Dame, Notre Dame, IN 46556}
\author{Thomas Parsons}
\affiliation{Department of Chemistry and Biochemistry, University
of Notre Dame, Notre Dame, IN 46556}
\author{Kathie E. Newman}
\affiliation{Department of Physics, University
of Notre Dame, Notre Dame, IN 46556}
\author{J. Daniel Gezelter}
\email{gezelter@nd.edu.}
\affiliation{Department of Chemistry and Biochemistry, University
of Notre Dame, Notre Dame, IN 46556}

\date{\today}% It is always \today, today,

\begin{abstract}
  In the first two papers in this series, we developed new shifted
  potential (SP), gradient shifted force (GSF), and Taylor shifted
  force (TSF) real-space methods for multipole interactions in
  condensed phase simulations.  Here, we discuss the dielectric
  properties of fluids that emerge from simulations using these
  methods.  Most electrostatic methods (including the Ewald sum)
  require correction to the conducting boundary fluctuation formula
  for the static dielectric constants, and we discuss the derivation
  of these corrections for the new real space methods. For quadrupolar
  fluids, the analogous material property is the quadrupolar
  susceptibility. As in the dipolar case, the fluctuation formula for
  the quadrupolar susceptibility has corrections that depend on the
  electrostatic method being utilized. One of the most important
  effects measured by both the static dielectric and quadrupolar
  susceptibility is the ability to screen charges embedded in the
  fluid.  We use potentials of mean force between solvated ions to
  discuss how geometric factors can lead to distance-dependent
  screening in both quadrupolar and dipolar fluids.
\end{abstract}

\maketitle

\section{Introduction}

Over the past several years, there has been increasing interest in
pairwise or ``real space'' methods for computing electrostatic
interactions in condensed phase
simulations.\cite{Wolf99,Zahn02,Kast03,Beckd05,Ma05,Wu:2005nr,Fennell06,Fukuda:2013uq,Stenqvist:2015ph,Wang:2016kx}
These techniques were initially developed by Wolf {\it et al.}  in
their work towards an $\mathcal{O}(N)$ Coulombic sum.\cite{Wolf99}
Wolf's method of using cutoff neutralization and electrostatic damping
is able to obtain excellent agreement with Madelung energies in ionic
crystals.\cite{Wolf99}

Zahn \textit{et al.}\cite{Zahn02} and Fennell and Gezelter extended
this method using shifted force approximations at the cutoff distance
in order to conserve total energy in molecular dynamics
simulations.\cite{Fennell06} Other recent advances in real-space
methods for systems of point charges have included explicit
elimination of the net multipole moments inside the cutoff sphere
around each charge site.\cite{Fukuda:2013uq,Wang:2016kx}

In the previous two papers in this series, we developed three
generalized real space methods: shifted potential (SP), gradient
shifted force (GSF), and Taylor shifted force (TSF).\cite{PaperI,
  PaperII} These methods evaluate electrostatic interactions for
charges and higher order multipoles using a finite-radius cutoff
sphere.  The neutralization and damping of local moments within the
cutoff sphere is a multipolar generalization of Wolf's sum.  In the
GSF and TSF methods, additional terms are added to the potential
energy so that forces and torques also vanish smoothly at the cutoff
radius.  This ensures that the total energy is conserved in molecular
dynamics simulations. 

One of the most stringent tests of any new electrostatic method is the
fidelity with which that method can reproduce the bulk-phase
polarizability or equivalently, the dielectric properties of a
fluid. Before the advent of computer simulations, Kirkwood and Onsager
developed fluctuation formulae for the dielectric properties of
dipolar fluids.\cite{Kirkwood39,Onsagar36} Along with projections of
the frequency-dependent dielectric to zero frequency, these
fluctuation formulae are now widely used to predict the static
dielectric constants of simulated materials.

If we consider a system of dipolar or quadrupolar molecules under the
influence of an external field or field gradient, the net polarization
of the system will largely be proportional to the applied
perturbation.\cite{Chitanvis96, Stern03, SalvchovI14, SalvchovII14} In
simulations, the net polarization of the system is also determined by
the interactions \textit{between} the molecules. Therefore the
macroscopic polarizability obtained from a simulation depends on the
details of the electrostatic interaction methods that were employed in
the simulation. To determine the relevant physical properties of the
multipolar fluid from the system fluctuations, the interactions
between molecules must be incorporated into the formalism for the bulk
properties.

In most simulations, bulk materials are treated using periodic
replicas of small regions, and this level of approximation requires
corrections to the fluctuation formulae that were derived for the bulk
fluids. In 1983 Neumann proposed a general formula for evaluating
dielectric properties of dipolar fluids using both Ewald and
real-space cutoff methods.\cite{NeumannI83} Steinhauser and Neumann
used this formula to evaluate the corrected dielectric constant for
the Stockmayer fluid using two different methods: Ewald-Kornfeld (EK)
and reaction field (RF) methods.\cite{NeumannII83}

Zahn \textit{et al.}\cite{Zahn02} utilized this approach and evaluated
the correction factor for using damped shifted charge-charge
kernel. This was later generalized by Izvekov \textit{et
  al.},\cite{Izvekov08} who noted that the expression for the
dielectric constant reduces to widely-used conducting boundary formula
for real-space methods that have first derivatives that vanish at the
cutoff radius.

One of the primary topics of this paper is the derivation of
correction factors for the three new real space methods. The
corrections are modifications to fluctuation expressions to account
for truncation, shifting, and damping of the field and field gradient
contributions from other multipoles. We find that the correction
formulae for dipolar molecules depends not only on the methodology
being used, but also on whether the molecular dipoles are treated
using point charges or point dipoles. We derive correction factors for
both cases.

In quadrupolar fluids, the relationship between quadrupolar
susceptibility and dielectric screening is not as straightforward as
in the dipolar case.  The effective dielectric constant depends on the
geometry of the external (or internal) field
perturbation.\cite{Ernst92} Significant efforts have been made to
increase our understanding the dielectric properties of these
fluids,\cite{JeonI03,JeonII03,Chitanvis96} although a general
correction formula has not yet been developed.

In this paper we derive general formulae for calculating the
quadrupolar susceptibility of quadrupolar fluids. We also evaluate the
correction factor for SP, GSF, and TSF methods for quadrupolar fluids
interacting via point charges, point dipoles or directly through
quadrupole-quadrupole interactions.

\begin{figure}
\includegraphics[width=\linewidth]{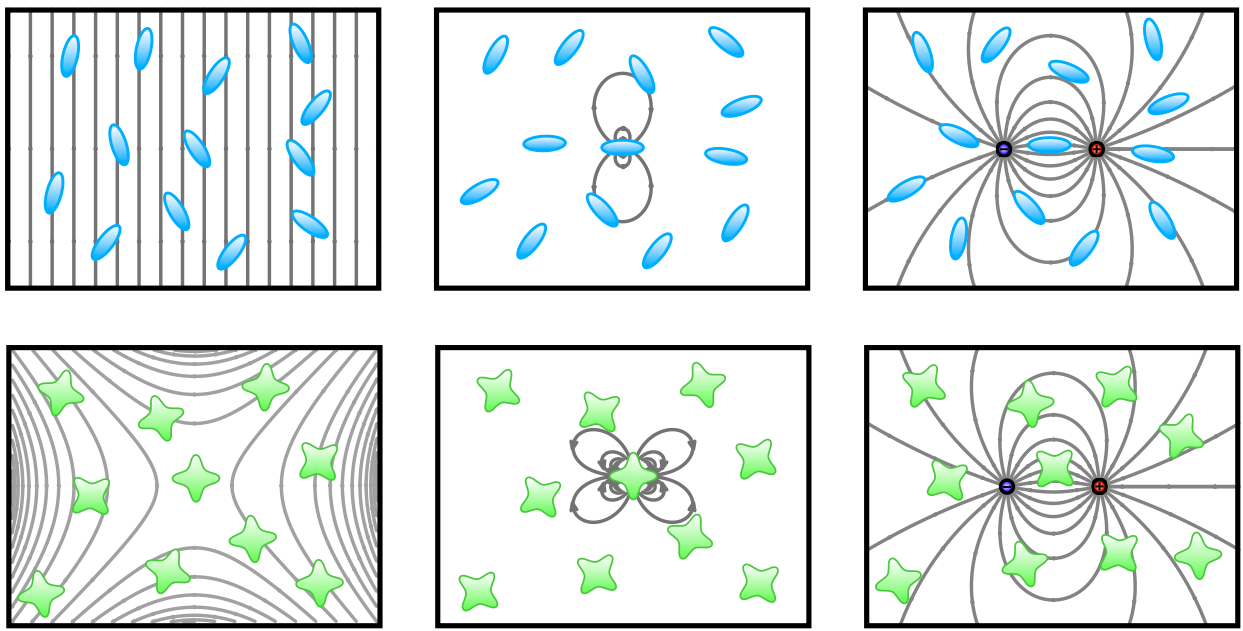}
\caption{Dielectric properties of a fluid measure the response to
  external electric fields and gradients (left), or internal fields
  and gradients generated by the molecules themselves (center), or
  fields produced by embedded ions (right). The dielectric constant
  ($\epsilon$) measures all three responses in dipolar fluids (top).
  In quadrupolar liquids (bottom), the relevant bulk property is the
  quadrupolar susceptibility ($\chi_Q$), and the geometry of the field
  determines the effective dielectric screening.}
\label{fig:schematic}
\end{figure}

We also calculate the screening behavior for two ions immersed in
multipolar fluids to estimate the distance dependence of charge
screening in both dipolar and quadrupolar fluids.  We use three
distinct methods to compare our analytical results with computer
simulations (see Fig. \ref{fig:schematic}):
\begin{enumerate}
\item responses of the fluid to external perturbations,
\item fluctuations of system multipole moments, and
\item potentials of mean force between solvated ions,
\end{enumerate}

Under the influence of weak external fields, the bulk polarization of
the system is primarily a linear response to the perturbation, where
the proportionality constant depends on the electrostatic interactions
between the multipoles. The fluctuation formulae connect bulk
properties of the fluid to equilibrium fluctuations in the system
multipolar moments during a simulation. These fluctuations also depend
on the form of the electrostatic interactions between molecules.
Therefore, the connections between the actual bulk properties and both
the computed fluctuation and external field responses must be modified
accordingly.

The potential of mean force (PMF) allows calculation of an effective
dielectric constant or screening factor from the potential energy
between ions before and after dielectric material is introduced.
Computing the PMF between embedded point charges is an additional
check on the bulk properties computed via the other two methods.

\section{The Real-Space Methods}

In the first paper in this series, we derived interaction energies, as
well as expressions for the forces and torques for point multipoles
interacting via three new real-space methods.\cite{PaperI} The Taylor
shifted-force (TSF) method modifies the electrostatic kernel,
$f(r) = 1/r$, so that all forces and torques go smoothly to zero at
the cutoff radius,
\begin{equation}
U^{\text{TSF}}_{ab} = M_{a} M_{b} f_n(r).
\label{TSF}
\end{equation}
Here the multipole operator for site $a$, $M_{a}$, is expressed in
terms of the point charge, $C_{a}$, dipole, ${\bf D}_{a}$, and
quadrupole, $\mathsf{Q}_{a}$, for object $a$, etc.  Because each of
the multipole operators includes gradient operators (one for a dipole,
two for a quadrupole, \textit{etc.}), an approximate electrostatic
kernel, $f_n(r)$ is Taylor-expanded around the cutoff radius, so that
$n + 1$ derivatives vanish as $r \rightarrow r_c$.  This ensures
smooth convergence of the energy, forces, and torques as molecules
leave and reenter each others cutoff spheres.  The order of the Taylor
expansion is determined by the multipolar order of the interaction.
That is, smooth quadrupole-quadrupole forces require the fifth
derivative to vanish at the cutoff radius, so the appropriate function
Taylor expansion will be of fifth order.

Following this procedure results in separate radial functions for each
of the distinct orientational contributions to the potential.  For
example, in dipole-dipole interactions, the direct dipole dot product
($\mathbf{D}_{a} \cdot \mathbf{D}_{b}$) is treated differently than
the dipole-distance dot products:
\begin{equation}
U_{\mathbf{D}_{a}\mathbf{D}_{b}}(r)= -\frac{1}{4\pi \epsilon_0} \left[ \left(
  \mathbf{D}_{a} \cdot
\mathbf{D}_{b} \right) v_{21}(r) +
\left( \mathbf{D}_{a} \cdot \hat{\mathbf{r}} \right)
\left( \mathbf{D}_{b} \cdot \hat{\mathbf{r}} \right) v_{22}(r) \right]
\end{equation}
In standard electrostatics, the two radial functions, $v_{21}(r)$ and
$v_{22}(r)$, are proportional to $1/r^3$, but they have distinct radial
dependence in the TSF method.  Careful choice of these functions makes
the forces and torques vanish smoothly as the molecules drift beyond
the cutoff radius (even when those molecules are in different
orientations).

A second and somewhat simpler approach involves shifting the gradient
of the Coulomb potential for each particular multipole order,
\begin{equation}
U^{\text{GSF}}_{ab} =  \sum \left[ U(\mathbf{r}, \mathsf{A}, \mathsf{B}) -
U(r_c \hat{\mathbf{r}},\mathsf{A}, \mathsf{B}) - (r-r_c)
\hat{\mathbf{r}} \cdot \nabla U(r_c \hat{\mathbf{r}},\mathsf{A},
\mathsf{B}) \right]
\label{generic2}
\end{equation}
where the sum describes a separate force-shifting that is applied to
each orientational contribution to the energy, i.e. $v_{21}$ and
$v_{22}$ are shifted separately. In this expression,
$\hat{\mathbf{r}}$ is the unit vector connecting the two multipoles
($a$ and $b$) in space, and $\mathsf{A}$ and $\mathsf{B}$ represent
the orientations of the multipoles.  Because this procedure is equivalent
to using the gradient of an image multipole placed at the cutoff
sphere for shifting the force, this method is called the gradient
shifted-force (GSF) approach.

Both the TSF and GSF approaches can be thought of as multipolar
extensions of the original damped shifted-force (DSF) approach that
was developed for point charges. There is also a multipolar extension
of the Wolf sum that is obtained by projecting an image multipole onto
the surface of the cutoff sphere, and including the interactions with
the central multipole and the image. This effectively shifts only the
total potential to zero at the cutoff radius,
\begin{equation}
U^{\text{SP}}_{ab} = \sum \left[ U(\mathbf{r}, \mathsf{A}, \mathsf{B}) -
U(r_c \hat{\mathbf{r}},\mathsf{A}, \mathsf{B}) \right]
\label{eq:SP}
\end{equation}          
where the sum again describes separate potential shifting that is done
for each orientational contribution to the energy.  The potential
energy between a central multipole and other multipolar sites goes
smoothly to zero as $r \rightarrow r_c$, but the forces and torques
obtained from this shifted potential (SP) approach are discontinuous
at $r_c$.

All three of the new real space methods share a common structure: the
various orientational contributions to multipolar interaction energies
require separate treatment of their radial functions, and these are
tabulated for both the raw Coulombic kernel ($1/r$) as well as the
damped kernel ($\mathrm{erfc}(\alpha r)/r$), in the first paper of this
series.\cite{PaperI} The second paper in this series evaluated the
fidelity with which the three new methods reproduced Ewald-based
results for a number of model systems.\cite{PaperII} One of the major
findings was that moderately-damped GSF simulations produced nearly
identical behavior with Ewald-based simulations, but the real-space
methods scale linearly with system size.

\section{Dipolar Fluids and the Dielectric Constant}

Dielectric properties of a fluid arise mainly from responses of the
fluid to either applied fields or transient fields internal to the
fluid. In response to an applied field, the molecules have electronic
polarizabilities, changes to internal bond lengths and angles, and
reorientations towards the direction of the applied field. There is an
added complication that in the presence of external field, the
perturbation experienced by any single molecule is not only due to
the external field but also to the fields produced by the all other
molecules in the system.

\subsection{Response to External Perturbations}

In the presence of uniform electric field $\mathbf{E}$, an individual
molecule with a permanent dipole moment $p_o$ will realign along the
direction of the field with an average polarization given by
\begin{equation}
\braket{\mathbf{p}} = \epsilon_0 \alpha_p \mathbf{E},
\end{equation}
where $\alpha_p = {p_o}^2 / 3 \epsilon_0 k_B T$ is the contribution to
molecular polarizability due solely to reorientation dynamics.
Because the applied field must overcome thermal motion, the
orientational polarization depends inversely on the temperature.

A condensed phase system of permanent dipoles will also polarize along
the direction of an applied field. The polarization density of the
system is
\begin{equation}
\textbf{P} = \epsilon_o \chi_{D} \mathbf{E},
\label{pertDipole}
\end{equation} 
where the constant $\chi_D$ is the dipole susceptibility, which is an
emergent property of the dipolar fluid, and is the quantity most
directly related to the static dielectric constant,
$\epsilon = 1 + \chi_D$.

\subsection{Fluctuation Formula}

For a system of dipolar molecules at thermal equilibrium, we can
define both a system dipole moment, $\mathbf{M} = \sum_i \mathbf{p}_i$
as well as a dipole polarization density,
$\mathbf{P} = \braket{\mathbf{M}}/V$.  The polarization density can be
expressed approximately in terms of fluctuations in the net dipole
moment,
\begin{equation}
\mathbf{P} = \epsilon_o \frac{\braket{\mathbf{M}^2}-{\braket{\mathbf{M}}}^2}{3 \epsilon_o V k_B T}\bf{E}
\label{flucDipole}
\end{equation}
This has structural similarity with the Boltzmann average for the
polarization of a single molecule. Here
$ \braket{\mathbf{M}^2}-{\braket{\mathbf{M}}}^2$ measures fluctuations
in the net dipole moment,
\begin{equation}
 \langle \mathbf{M}^2 \rangle - \langle \mathbf{M} \rangle^2 =
 \langle M_x^2+M_y^2+M_z^2 \rangle - \left( \langle M_x \rangle^2 +
   \langle M_y \rangle^2 +
   \langle M_z \rangle^2 \right).
\label{eq:flucDip}
\end{equation}
When no applied electric field is present, the ensemble average of
both the net dipole moment $\braket{\mathbf{M}}$ and dipolar
polarization $\bf{P}$ tends to vanish but $\braket{\mathbf{M}^2}$ does
not. The bulk dipole polarizability can therefore be
written
\begin{equation}
  \alpha_D = \frac{\braket{\mathbf{M}^2}-{\braket{\mathbf{M}}}^2}{3 \epsilon_o V k_B T}.
\end{equation}
The susceptibility ($\chi_D$) and bulk polarizability ($\alpha_D$)
both measure responses of a dipolar system.  However, $\chi_D$ is the
bulk property assuming an infinite system and exact treatment of
electrostatic interactions, while $\alpha_D$ is relatively simple to
compute from numerical simulations.  One of the primary aims of this
paper is to provide the connection between the bulk properties
($\epsilon, \chi_D$) and the computed quantities ($\alpha_D$) that
have been adapted for the new real-space methods.

\subsection{Correction Factors}
\label{sec:corrFactor}
In the presence of a uniform external field $ \mathbf{E}^\circ$, the
total electric field at $\mathbf{r}$ depends on the polarization
density at all other points in the system,\cite{NeumannI83}
\begin{equation}
\mathbf{E}(\mathbf{r}) = \mathbf{E}^\circ(\mathbf{r}) +
\frac{1}{4\pi\epsilon_o} \int d\mathbf{r}^\prime
\mathbf{T}(\mathbf{r}-\mathbf{r}^\prime)\cdot
{\mathbf{P}(\mathbf{r}^\prime)}.
\label{eq:localField}
\end{equation}
$\mathbf{T}$ is the dipole interaction tensor connecting dipoles at
$\mathbf{r}^\prime$ with the point of interest ($\mathbf{r}$), where
the integral is done over all space.  Because simulations utilize
periodic boundary conditions or spherical cutoffs, the integral is
normally carried out either over the domain ($0 < r < r_c$) or in
reciprocal space.

\begin{figure}
\includegraphics[width=\linewidth]{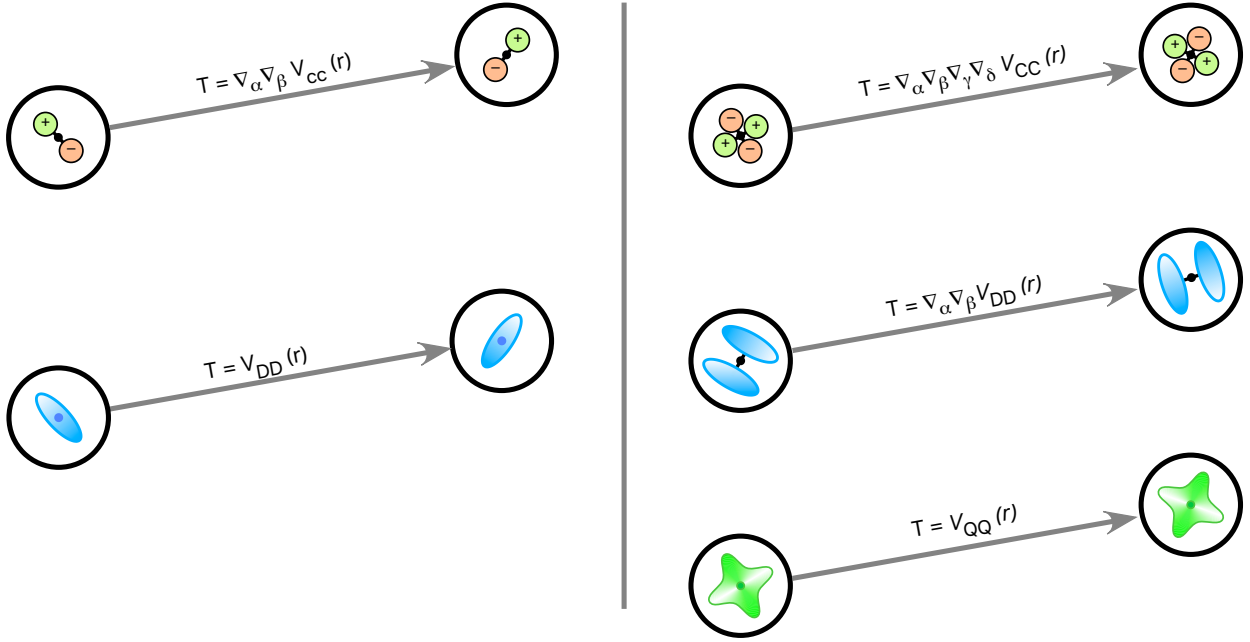}
\caption{In the real-space electrostatic methods, the molecular dipole
  tensor, $\mathbf{T}_{\alpha\beta}(r)$, is not the same for
  charge-charge interactions as for point dipoles (left panel). The
  same holds true for the molecular quadrupole tensor (right panel),
  $\mathbf{T}_{\alpha\beta\gamma\delta}(r)$, which can have distinct
  forms if the molecule is represented by charges, dipoles, or point
  quadrupoles.}
\label{fig:tensor}
\end{figure}

In simulations of dipolar fluids, the molecular dipoles may be
represented either by closely-spaced point charges or by 
point dipoles (see Fig. \ref{fig:tensor}).
In the case where point charges are interacting via an electrostatic
kernel, $v(r)$, the effective {\it molecular} dipole tensor,
$\mathbf{T}$ is obtained from two successive applications of the
gradient operator to the electrostatic kernel,
\begin{eqnarray}
T_{\alpha \beta}(\mathbf{r}) &=&  \nabla_\alpha \nabla_\beta
                               \left(v(r)\right)  \\
  &=& \delta_{\alpha \beta}
\left(\frac{1}{r} v^\prime(r) \right) + \frac{r_{\alpha}
  r_{\beta}}{r^2} \left( v^{\prime \prime}(r) - \frac{1}{r}
  v^{\prime}(r) \right)
\label{dipole-chargeTensor}
\end{eqnarray}
where $v(r)$ may be either the bare kernel ($1/r$) or one of the
modified (Wolf or DSF) kernels.  This tensor describes the effective
interaction between molecular dipoles ($\mathbf{D}$) in Gaussian units
as $-\mathbf{D} \cdot \mathbf{T} \cdot \mathbf{D}$.

When utilizing any of the three new real-space methods for point
\textit{dipoles}, the tensor is explicitly constructed,
\begin{equation}
T_{\alpha \beta}(\mathbf{r})  =  \delta_{\alpha \beta} v_{21}(r) +
\frac{r_{\alpha} r_{\beta}}{r^2} v_{22}(r) 
\label{dipole-diopleTensor}
\end{equation}
where the functions $v_{21}(r)$ and $v_{22}(r)$ depend on the level of
the approximation.\cite{PaperI,PaperII} Although the Taylor-shifted
(TSF) and gradient-shifted (GSF) models produce to the same $v(r)$
function for point charges, they have distinct forms for the
dipole-dipole interaction.
 
Using the constitutive relation in Eq. (\ref{pertDipole}), the
polarization density $\mathbf{P}(\mathbf{r})$ is given by,
\begin{equation}
\mathbf{P}(\mathbf{r}) = \epsilon_o \chi_D
\left(\mathbf{E}^\circ(\mathbf{r}) + \frac{1}{4\pi\epsilon_o} \int
  d\mathbf{r}^\prime \mathbf{T}(\mathbf{r}-\mathbf{r}^\prime ) \cdot \mathbf{P}(\mathbf{r}^\prime)\right).
\label{constDipole}
\end{equation}
Note that $\chi_D$ depends explicitly on the details of the dipole
interaction tensor.  Neumann \textit{et al.}
\cite{NeumannI83,NeumannII83,Neumann84,Neumann85} derived an elegant
way to modify the fluctuation formula to correct for approximate
interaction tensors. This correction was derived using a Fourier
representation of the interaction tensor,
$\tilde{\mathbf{T}}(\mathbf{k})$, and involves the quantity,
\begin{equation}
  \mathsf{A} = \frac{3}{4\pi}\tilde{\mathbf{T}}(0) = \frac{3}{4\pi} \int_V
  d\mathbf{r} \mathbf{T}(\mathbf{r})
\end{equation}
which is the $k \rightarrow 0$ limit of
$\tilde{\mathbf{T}}(\mathbf{k})$.  Note that the integration of the
dipole tensors, Eqs. (\ref{dipole-chargeTensor}) and
(\ref{dipole-diopleTensor}), over spherical volumes yields values only
along the diagonal.  Additionally, the spherical symmetry of
$\mathbf{T}(\mathbf{r})$ insures that all diagonal elements are
identical.  For this reason, $\mathsf{A}$ can be written as a scalar
constant ($A$) multiplying the unit tensor.

Using the quantity $A$ (originally called $Q$ in
refs. \onlinecite{NeumannI83,NeumannII83,Neumann84,Neumann85}), the
dielectric constant can be computed
\begin{equation}
\epsilon = \frac{3+(A + 2)(\epsilon_{CB}-1)}{3 + (A -1)(\epsilon_{CB}-1)}
\label{correctionFormula}
\end{equation}
where $\epsilon_{CB}$ is the widely-used conducting boundary
expression for the dielectric constant,
\begin{equation}
\epsilon_{CB} = 1 + \frac{\braket{\bf{M}^2}-{\braket{\bf{M}}}^2}{3
  \epsilon_o V k_B T} = 1 + \alpha_{D}.
\label{conductingBoundary}
\end{equation}
Eqs. (\ref{correctionFormula}) and (\ref{conductingBoundary}) allow
estimation of the static dielectric constant from fluctuations
computed directly from simulations, with the understanding that
Eq. (\ref{correctionFormula}) is extraordinarily sensitive when $A$ is
far from unity.

We have utilized the Neumann \textit{et al.} approach for the three
new real-space methods, and obtain method-dependent correction
factors.  The expression for the correction factor also depends on
whether the simulation involves point charges or point dipoles to
represent the molecular dipoles.  These corrections factors are listed
in Table \ref{tab:A}.  We note that the GSF correction factor for
point dipoles has been independently derived by Stenqvist \textit{et
  al.}\cite{Stenqvist:2015ph}
\begin{table}
  \caption{Expressions for the dipolar correction factor ($A$) for the
    real-space electrostatic methods in terms of the damping parameter 
    ($\alpha$) and the cutoff radius ($r_c$).  The Ewald-Kornfeld result 
    derived in Refs. \onlinecite{Adams80,Adams81,NeumannI83} is shown
    for comparison using the Ewald convergence parameter ($\kappa$)
    and the real-space cutoff value ($r_c$). } 
\label{tab:A}
\begin{tabular}{l|c|c}
\toprule       
& \multicolumn{2}{c}{Molecular Representation} \\ 
Method & point charges & point dipoles  \\
\colrule
 Shifted Potential (SP) & $ \mathrm{erf}(r_c \alpha) - \frac{2
                               \alpha r_c}{\sqrt{\pi}} e^{-\alpha^2
                               r_c^2}$ & $\mathrm{erf}(r_c \alpha)
                                         -\frac{2 \alpha
                                         r_c}{\sqrt{\pi}}\left(1+\frac{2\alpha^2
                                         {r_c}^2}{3}
                                         \right)e^{-\alpha^2{r_c}^2}
                                         $\\ \colrule
Gradient-shifted  (GSF) & 1 & $\mathrm{erf}(\alpha  r_c)-\frac{2
                              \alpha  r_c}{\sqrt{\pi}}  \left(1 +
                              \frac{2 \alpha^2 r_c^2}{3} +
                              \frac{\alpha^4
                              r_c^4}{3}\right)e^{-\alpha^2 r_c^2} $ \\ \colrule
 Taylor-shifted  (TSF) &\multicolumn{2}{c}{1}  \\ \colrule
%Spherical Cutoff (SC)& \multicolumn{2}{c}{$\mathrm{erf}(r_c \alpha) -
%                        \frac{2 \alpha r_c}{\sqrt{\pi}} e^{-\alpha^2
%                        r_c^2}$} \\ 
Ewald-Kornfeld (EK) & \multicolumn{2}{c}{$\mathrm{erf}(r_c \kappa) -
                      \frac{2 \kappa r_c}{\sqrt{\pi}} e^{-\kappa^2
                      r_c^2}$}  \\
\botrule
\end{tabular}
\end{table}
Note that for point charges, the GSF and TSF methods produce estimates
of the dielectric that need no correction, and the TSF method likewise
needs no correction for point dipoles. 

\section{Quadrupolar Fluids and the Quadrupolar Susceptibility}
\subsection{Response to External Perturbations}

A molecule with a permanent quadrupole, $\mathsf{q}$, will align in
the presence of an electric field gradient $\nabla\mathbf{E}$.  The
anisotropic polarization of the quadrupole is given
by,\cite{AduGyamfi78,AduGyamfi81}
\begin{equation}
\braket{\mathsf{q}} - \frac{\mathbf{I}}{3}
\mathrm{Tr}(\mathsf{q}) = \epsilon_o \alpha_q \nabla\mathbf{E},
\end{equation}
where $\alpha_q = q_o^2 / 15 \epsilon_o k_B T $ is a molecular quadrupole
polarizability and $q_o$ is an effective quadrupole moment for the molecule,
\begin{equation}
 q_o^2 = 3 \mathsf{q}:\mathsf{q} - \mathrm{Tr}(\mathsf{q})^2.
\end{equation}
Note that quadrupole calculations involve tensor contractions (double
dot products) between rank two tensors, which are defined as
\begin{equation}
 \mathsf{A} \colon \mathsf{B} = \sum_\alpha \sum_\beta
 A_{\alpha \beta} B_{\beta \alpha}.
\end{equation}

In the presence of an external field gradient, a system of quadrupolar
molecules also organizes with an anisotropic polarization,
\begin{equation}
\mathsf{Q} - \frac{\mathbf{I}}{3} \mathrm{Tr}(\mathsf{Q}) =  \epsilon_o
\chi_Q  \nabla\mathbf{E}
\end{equation}
where $\mathsf{Q}$ is the traced quadrupole density of the system and
$\chi_Q$ is a macroscopic quadrupole susceptibility which has
dimensions of $\mathrm{length}^{-2}$. Equivalently, the traceless form
may be used,
\begin{equation}
\mathsf{\Theta} = 3 \epsilon_o  \chi_Q \nabla\mathbf{E},
\label{pertQuad}
\end{equation} 
where
$\mathsf{\Theta} = 3\mathsf{Q} - \mathbf{I} \mathrm{Tr}(\mathsf{Q})$
is the traceless tensor that also describes the system quadrupole
density.  It is this tensor that will be utilized to derive correction
factors below.

\subsection{Fluctuation Formula}
As in the dipolar case, we may define a system quadrupole moment,
$\mathsf{M}_Q = \sum_i \mathsf{q}_i$ and the traced quadrupolar
density, $\mathsf{Q} = \mathsf{M}_Q / V$.  A fluctuation formula can
be written for a system comprising quadrupolar
molecules,\cite{LoganI81,LoganII82,LoganIII82}
\begin{equation}
\mathsf{Q} - \frac{\mathbf{I}}{3} \mathrm{Tr}(\mathsf{Q}) = \epsilon_o
\frac{\braket{\mathsf{M}_Q^2}-{\braket{\mathsf{M}_Q}}^2}{15 \epsilon_o
  V k_B T} \nabla\mathbf{E}.
\label{flucQuad}
\end{equation}
Some care is needed in the definitions of the averaged quantities.  These
refer to the effective quadrupole moment of the system, and they are
computed as follows,
\begin{align}
\braket{\mathsf{M}_Q^2} &= \braket{3 \mathsf{M}_Q:\mathsf{M}_Q -
  \mathrm{Tr}(\mathsf{M}_Q)^2 }\\
\braket{\mathsf{M}_Q}^2 &= 3 \braket{\mathsf{M}_Q}:\braket{\mathsf{M}_Q} -
\mathrm{Tr}(\braket{\mathsf{M}_Q})^2
\label{eq:flucQuad}
\end{align}
The bulk quadrupolarizability is given by,
\begin{equation}
  \alpha_Q = \frac{\braket{\mathsf{M}_Q^2}-{\braket{\mathsf{M}_Q}}^2}{15 \epsilon_o V k_B T}.
\label{propConstQuad}
\end{equation}
Note that as in the dipolar case, $\alpha_Q$ and $\chi_Q$ are distinct
quantities.  $\chi_Q$ measures the bulk response assuming an infinite
system and exact electrostatics, while $\alpha_Q$ is relatively simple
to compute from numerical simulations.  As in the dipolar case,
estimation of the true bulk property requires correction for
truncation, shifting, and damping of the electrostatic interactions.

\subsection{Correction Factors}
In this section we generalize the treatment of Neumann \textit{et al.}
for quadrupolar fluids. Interactions involving multiple quadrupoles
are rank 4 tensors, and we therefore describe quantities in this
section using Einstein notation.

In the presence of a uniform external field gradient,
$\partial_\alpha {E}^\circ_\beta $, the total field gradient at
$\mathbf{r}$ depends on the quadrupole polarization density at all
other points in the system,
\begin{equation}
\partial_\alpha E_\beta(\mathbf{r}) = \partial_\alpha
{E}^\circ_\beta(\mathbf{r}) + \frac{1}{8\pi \epsilon_o}\int
T_{\alpha\beta\gamma\delta}({\mathbf{r}-\mathbf{r}^\prime})
Q_{\gamma\delta}(\mathbf{r}^\prime) d\mathbf{r}^\prime 
\label{gradMaxwell}
\end{equation}
where and $T_{\alpha\beta\gamma\delta}$ is the quadrupole interaction
tensor connecting quadrupoles at $\mathbf{r}^\prime$ with the point of
interest ($\mathbf{r}$).

% \begin{figure}
% \includegraphics[width=3in]{QuadrupoleFigure}
% \caption{With the real-space electrostatic methods, the molecular
%   quadrupole tensor, $\mathbf{T}_{\alpha\beta\gamma\delta}(r)$,
%   governing interactions between molecules is not the same for
%   quadrupoles represented via sets of charges, point dipoles, or by
%   single point quadrupoles.}
% \label{fig:quadrupolarFluid}
% \end{figure}

In simulations of quadrupolar fluids, the molecular quadrupoles may be
represented by closely-spaced point charges, by multiple point
dipoles, or by a single point quadrupole (see
Fig. \ref{fig:tensor}).  In the case where point charges are
interacting via an electrostatic kernel, $v(r)$, the effective
molecular quadrupole tensor can obtained from four successive
applications of the gradient operator to the electrostatic kernel,
\begin{eqnarray}
T_{\alpha\beta\gamma\delta}(\mathbf{r}) &=& \nabla_\alpha \nabla_\beta
                                   \nabla_\gamma \nabla_\delta v(r) \\
 &=& \left(\delta_{\alpha\beta}\delta_{\gamma\delta} +
     \delta_{\alpha\gamma}\delta_{\beta\delta}+
     \delta_{\alpha\delta}\delta_{\beta\gamma}\right)\left(-\frac{v^\prime(r)}{r^3}
     + \frac{v^{\prime \prime}(r)}{r^2}\right) \nonumber \\
 & &+ \left(\delta_{\alpha\beta} r_\gamma r_\delta + 5  \mathrm{~permutations}
     \right) \left(\frac{3v^\prime(r)}{r^5}-\frac{3v^{\prime \prime}(r)}{r^4} +
     \frac{v^{\prime \prime \prime}(r)}{r^3}\right) \nonumber \\
 & &+ r_\alpha r_\beta r_\gamma r_\delta
     \left(-\frac{15v^\prime(r)}{r^7}+\frac{15v^{\prime \prime}(r)}{r^6}-\frac{6v^{\prime
     \prime \prime}(r)}{r^5} + \frac{v^{\prime \prime \prime \prime}(r)}{r^4}\right),
\label{quadCharge}
\end{eqnarray}
where $v(r)$ can either be the electrostatic kernel ($1/r$) or one of
the modified (Wolf or DSF) kernels.  

Similarly, when representing quadrupolar molecules with multiple point
\textit{dipoles}, the molecular quadrupole interaction tensor can be
obtained using two successive applications of the gradient operator to
the dipole interaction tensor,
\begin{eqnarray}
T_{\alpha\beta\gamma\delta}(\mathbf{r}) &=& \nabla_\alpha \nabla_\beta
                                            T_{\gamma\delta}(\mathbf{r}) \\ 
& = & \delta_{\alpha\beta}\delta_{\gamma\delta} \frac{v^\prime_{21}(r)}{r} +
      \left(\delta_{\alpha\gamma}\delta_{\beta\delta}+
      \delta_{\alpha\delta}\delta_{\beta\gamma}\right)\frac{v_{22}(r)}{r^2}
      \nonumber\\ 
 & &+ \delta_{\gamma\delta} r_\alpha r_\beta
     \left(\frac{v^{\prime \prime}_{21}(r)}{r^2}-\frac{v^\prime_{21}(r)}{r^3} \right)\nonumber \\
 & &+\left(\delta_{\alpha\beta} r_\gamma r_\delta +
     \delta_{\alpha\gamma} r_\beta r_\delta  +\delta_{\alpha\delta}
     r_\gamma r_\beta + \delta_{\beta\gamma} r_\alpha r_\delta
     +\delta_{\beta\delta} r_\alpha r_\gamma  \right)
     \left(\frac{v^\prime_{22}(r)}{r^3}-\frac{2v_{22}(r)}{r^4}\right)
     \nonumber \\
 & &+ r_\alpha r_\beta r_\gamma r_\delta
     \left(\frac{v^{\prime
     \prime}_{22}(r)}{r^4}-\frac{5v^\prime_{22}(r)}{r^5}+\frac{8v_{22}(r)}{r^6}\right),
\label{quadDip}
\end{eqnarray}
where $T_{\gamma\delta}(\mathbf{r})$ is a dipole-dipole interaction
tensor that depends on the level of the approximation (see
Eq. (\ref{dipole-diopleTensor})).\cite{PaperI,PaperII} Similarly
$v_{21}(r)$ and $v_{22}(r)$ are the radial functions for different real
space cutoff methods defined in the first paper in this
series.\cite{PaperI}

For quadrupolar liquids modeled using point quadrupoles, the
interaction tensor can be constructed as,
\begin{eqnarray}
T_{\alpha\beta\gamma\delta}(\mathbf{r}) &=&
                                              \left(\delta_{\alpha\beta}\delta_{\gamma\delta}
                                              +
                                              \delta_{\alpha\gamma}\delta_{\beta\delta}+
                                              \delta_{\alpha\delta}\delta_{\beta\gamma}\right)v_{41}(r)
                                              + (\delta_{\gamma\delta} r_\alpha r_\beta +  \mathrm{ 5\; permutations}) \frac{v_{42}(r)}{r^2} \nonumber \\  
& & + r_\alpha r_\beta r_\gamma r_\delta  \left(\frac{v_{43}(r)}{r^4}\right), 
\label{quadRadial}
\end{eqnarray}
where again $v_{41}(r)$, $v_{42}(r)$, and $v_{43}(r)$ are radial
functions defined in Paper I of the series. \cite{PaperI} Note that
these radial functions have different functional forms depending on
the level of approximation being employed.

The integral in Eq. (\ref{gradMaxwell}) can be divided into two
parts, $|\mathbf{r}-\mathbf{r}^\prime|\rightarrow 0 $ and
$|\mathbf{r}-\mathbf{r}^\prime|> 0$. Since the self-contribution to
the field gradient vanishes at the singularity (see the supplemental
material), Eq. (\ref{gradMaxwell}) can be
written as,
\begin{equation}
\partial_\alpha E_\beta(\mathbf{r}) = \partial_\alpha {E}^\circ_\beta(\mathbf{r}) +
  \frac{1}{8\pi \epsilon_o}\int\limits_{|\mathbf{r}-\mathbf{r}^\prime|> 0 }
  T_{\alpha\beta\gamma\delta}(\mathbf{r}-\mathbf{r}^\prime)
  {Q}_{\gamma\delta}(\mathbf{r}^\prime) d\mathbf{r}^\prime.
\end{equation}
If $\mathbf{r} = \mathbf{r}^\prime$ is excluded from the integration,
the total gradient can be most easily expressed in terms of
traceless quadrupole density as below,\cite{LoganI81}
\begin{equation}
\partial_\alpha E_\beta(\mathbf{r}) = \partial_\alpha
{E}^\circ_\beta(\mathbf{r}) + \frac{1}{24\pi
  \epsilon_o}\int\limits_{|\mathbf{r}-\mathbf{r}^\prime|> 0 }
T_{\alpha\beta\gamma\delta}(\mathbf{r}-\mathbf{r}^\prime) \Theta_{\gamma\delta}(\mathbf{r}') d\mathbf{r}^\prime,
\end{equation}
where
$\Theta_{\alpha\beta} = 3Q_{\alpha\beta} - \delta_{\alpha\beta}Tr(Q)$
is the traceless quadrupole density. In analogy to
Eq. (\ref{pertQuad}) above, the quadrupole polarization density may
now be related to the quadrupolar susceptibility, $\chi_Q$,
\begin{equation}
\frac{1}{3}{\Theta}_{\alpha\beta}(\mathbf{r}) = \epsilon_o {\chi}_Q
\left[\partial_\alpha {E}^\circ_\beta(\mathbf{r}) + \frac{1}{24\pi
    \epsilon_o}\int\limits_{|\mathbf{r}-\mathbf{r}^\prime|> 0 }
  T_{\alpha\beta\gamma\delta}(\mathbf{r}-\mathbf{r}^\prime)
  \Theta_{\gamma\delta}(\mathbf{r}^\prime) d\mathbf{r}^\prime \right].
\end{equation}
For periodic boundaries and with a uniform imposed
$\partial_\alpha E^\circ_\beta$, the quadrupole density
${\Theta}_{\alpha\beta}$ will be uniform over the entire space. After
performing a Fourier transform (see the Appendix in
ref. \onlinecite{NeumannI83}) we obtain,
\begin{equation}
\frac{1}{3}\tilde{\Theta}_{\alpha\beta}(\mathbf{k})=
\epsilon_o {\chi}_Q \left[{\partial_\alpha
    \tilde{E}^\circ_\beta}(\mathbf{k})+ \frac{1}{24\pi
    \epsilon_o} \tilde{T}_{\alpha\beta\gamma\delta}(\mathbf{k})
 \tilde{\Theta}_{\gamma\delta}(\mathbf{k})\right].
\label{fourierQuad}
\end{equation} 
If the applied field gradient is homogeneous over the entire volume,
${\partial_ \alpha \tilde{E}^\circ_\beta}(\mathbf{k}) = 0 $ except at
$ \mathbf{k} = 0$. Similarly, the quadrupolar polarization density can
also considered uniform over entire space. As in the dipolar case,
\cite{NeumannI83} the only relevant contribution from the interaction
tensor will also be when $\mathbf{k} = 0$. Therefore Eq.
(\ref{fourierQuad}) can be written as,
\begin{equation}
\frac{1}{3}\tilde{\Theta}_{\alpha\beta}(\mathrm{0})=
\epsilon_o {\chi}_Q \left[{\partial_\alpha
    \tilde{E}^\circ_\beta}(\mathrm{0})+ \frac{1}{24\pi
    \epsilon_o} \tilde{T}_{\alpha\beta\gamma\delta}(\mathrm{0})
 \tilde{\Theta}_{\gamma\delta}(\mathrm{0})\right].
\label{fourierQuadZeroK}
\end{equation} 
The quadrupolar tensor
$\tilde{T}_{\alpha\beta\gamma\delta}(\mathrm{0})$ is a rank 4 tensor
with 81 elements. The only non-zero elements, however, are those with
two doubly-repeated indices, \textit{i.e.}
$\tilde{T}_{aabb}(\mathrm{0})$ and all permutations of these indices.
The special case of quadruply-repeated indices,
$\tilde{T}_{aaaa}(\mathrm{0})$ also survives (see appendix
\ref{ap:quadContraction}). Furthermore, for the both diagonal and
non-diagonal components of the quadrupolar polarization
$\tilde{\Theta}_{\alpha\beta}$, we can contract the second term in
Eq. \ref{fourierQuadZeroK} (see appendix
\ref{ap:quadContraction}):
\begin{equation}
\tilde{T}_{\alpha\beta\gamma\delta}(\mathrm{0})\tilde{\Theta}_{\gamma\delta}(\mathrm{0})=
8 \pi \mathrm{B} \tilde{\Theta}_{\alpha\beta}(\mathrm{0}).
\label{quadContraction}
\end{equation}
Here $\mathrm{B} = \tilde{T}_{abab}(\mathrm{0}) / 4 \pi$ for
$a \neq b$.  Using this quadrupolar contraction we can solve Eq.
\ref{fourierQuadZeroK} as follows
\begin{eqnarray}
\frac{1}{3}\tilde{\Theta}_{\alpha\beta}(\mathrm{0}) &=& \epsilon_o
                                                      {\chi}_Q
                                                      \left[{\partial_\alpha
                                                      \tilde{E}^\circ_\beta}(\mathrm{0})+
                                                      \frac{\mathrm{B}}{3
                                                      \epsilon_o} 
                                                      {\tilde{\Theta}}_{\alpha\beta}(\mathrm{0})\right]
                                                      \nonumber \\                                                    
&=& \left[\frac{\epsilon_o {\chi}_Q} {1-{\chi}_Q \mathrm{B}}\right]
{\partial_\alpha \tilde{E}^\circ_\beta}(\mathrm{0}).
\label{fourierQuad2}
\end{eqnarray}
In real space, the correction factor is found to be,
\begin{equation}
\mathrm{B} = \frac{1}{4 \pi} \tilde{T}_{abab}(0) = \frac{1}{4 \pi} \int_V {T}_{abab}(\mathbf{r}) d\mathbf{r},
\end{equation}
%If the applied field gradient is homogeneous over the
%entire volume, ${\partial_ \alpha \tilde{E}^\circ_\beta}(\mathbf{k}) = 0 $ except at
%$ \mathbf{k} = 0$.  As in the dipolar case, the only relevant
%contribution of the interaction tensor will also be when $\mathbf{k} = 0$.
%Therefore equation (\ref{fourierQuad}) can be written as,
%\begin{equation}
%\frac{1}{3}{\tilde{\Theta}}_{\alpha\beta}({0}) = \epsilon_o {\chi}_Q
%\left(1-\frac{1}{4\pi} {\chi}_Q \tilde{T}_{\alpha\beta\alpha\beta}({0})\right)^{-1} \partial_\alpha \tilde{E}^\circ_\beta({0})
%\label{fourierQuad2}
%\end{equation}
%where $\tilde{T}_{\alpha\beta\alpha\beta}({0})$ can be evaluated as,
%\begin{equation}
%\tilde{T}_{\alpha\beta\alpha\beta}({0}) = \int_V {T}_{\alpha\beta\alpha\beta}(\mathbf{r})d\mathbf{r}
%\label{realTensorQaud}
%\end{equation}
%We may now define a quantity to represent the contribution from the
%$\mathbf{k} = 0$ portion of the interaction tensor,
%\begin{equation}
%B = \frac{1}{4 \pi} \int_V T_{\alpha \beta \alpha \beta}(\mathbf{r})
%d\mathbf{r}
%\end{equation}
which has been integrated over the interaction volume $V$ and has
units of $\mathrm{length}^{-2}$.

In terms of the traced quadrupole moment, Eq. (\ref{fourierQuad2})
can be written,
\begin{equation}
\mathsf{Q} - \frac{\mathbf{I}}{3} \mathrm{Tr}(\mathsf{Q})
= \frac{\epsilon_o {\chi}_Q}{1-  {\chi}_Q \mathrm{B}} \nabla \mathbf{E}^\circ
\label{tracedConstQuad}
\end{equation}
Comparing (\ref{tracedConstQuad}) and (\ref{flucQuad}) we obtain,
\begin{equation}
\frac{\braket{\mathsf{M}_Q^2}-{\braket{\mathsf{M}_Q}}^2}{15 \epsilon_o
  V k_B T} = \frac{\chi_Q} {1 - \chi_Q \mathrm{B}},
\end{equation}
or equivalently,
\begin{equation}
\chi_Q = \frac{\braket{\mathsf{M}_Q^2}-{\braket{\mathsf{M}_Q}}^2}{15 \epsilon_o
  V k_B T} \left(1 + \mathrm{B} \frac{\braket{\mathsf{M}_Q^2}-{\braket{\mathsf{M}_Q}}^2}{15 \epsilon_o
  V k_B T} \right)^{-1}.
\label{eq:finalForm}
\end{equation}
Eq. (\ref{eq:finalForm}) now expresses a bulk property (the
quadrupolar susceptibility, $\chi_Q$) in terms of a fluctuation in the
system quadrupole moment and a quadrupolar correction factor
($\mathrm{B}$).  The correction factors depend on the cutoff method
being employed in the simulation, and these are listed in Table
\ref{tab:B}.

In terms of the macroscopic quadrupole polarizability, $\alpha_Q$,
which may be thought of as the ``conducting boundary'' version of the
susceptibility,
\begin{equation}
\chi_Q = \frac{\alpha_Q}{1 + \mathrm{B} \alpha_Q}.
\label{eq:quadrupolarSusceptiblity}
\end{equation}
If an electrostatic method produces $\mathrm{B} \rightarrow 0$, the computed
quadrupole polarizability and quadrupole susceptibility converge to
the same value.

\begin{sidewaystable}
  \caption{Expressions for the quadrupolar correction factor
    ($\mathrm{B}$) for the real-space electrostatic methods in terms
    of the damping parameter ($\alpha$) and the cutoff radius
    ($r_c$). The units of the correction factor are
    $ \mathrm{length}^{-2}$ for quadrupolar fluids.}
\label{tab:B}
\begin{tabular}{l|c|c|c} \toprule
\multirow{2}{*}{Method} &  \multicolumn{3}{c}{Molecular Representation} \\ 
 & charges & dipoles & quadrupoles \\\colrule
Shifted Potential (SP) & $ -\frac{8 \alpha^5 {r_c}^3e^{-\alpha^2 r_c^2}}{15\sqrt{\pi}} $ &  $-  \frac{3 \mathrm{erfc(r_c\alpha)}}{5{r_c}^2}- \frac{2 \alpha e^{-\alpha^2 r_c^2}(9+6\alpha^2 r_c^2+4\alpha^4 r_c^4)}{15{\sqrt{\pi}r_c}}$& $ -\frac{16 \alpha^7 {r_c}^5 e^{-\alpha^2 r_c^2                                 }}{45\sqrt{\pi}}$  \\
Gradient-shifted  (GSF) & $- \frac{8 \alpha^5 {r_c}^3e^{-\alpha^2 r_c^2}}{15\sqrt{\pi}} $ & 0 &  $-\frac{4{\alpha}^7{r_c}^5 e^{-\alpha^2 r_c^2}(-1+2\alpha ^2 r_c^2)}{45\sqrt{\pi}}$\\
Taylor-shifted  (TSF) &  $ -\frac{8 \alpha^5 {r_c}^3e^{-\alpha^2 r_c^2}}{15\sqrt{\pi}} $ & $\frac{4\;\mathrm{erfc(\alpha r_c)}}{{5r_c}^2} + \frac{8\alpha e^{-\alpha^2{r_c}^2}\left(3+ 2\alpha^2 {r_c}^2 +\alpha^4{r_c}^4 \right)}{15\sqrt{\pi}r_c}$ & $\frac{2\;\mathrm{erfc}(\alpha r_c )}{{r_c}^2} + \frac{4{\alpha}e^{-\alpha^2 r_c^2}\left(45 + 30\alpha ^2 {r_c}^2 + 12\alpha^4 {r_c}^4 + 3\alpha^6 {r_c}^6 + 2 \alpha^8 {r_c}^8\right)}{45\sqrt{\pi}{r_c}}$ \\
\colrule
%Spherical Cutoff (SC) & \multicolumn{3}{c}{$ -\frac{8 \alpha^5
%                        {r_c}^3e^{-\alpha^2 r_c^2}}{15\sqrt{\pi}} $}\\ 
Ewald-Kornfeld (EK) & \multicolumn{3}{c}{$ -\frac{8 \kappa^5 {r_c}^3e^{-\kappa^2 r_c^2}}{15\sqrt{\pi}}$} \\
\botrule
\end{tabular}
\end{sidewaystable}

\section{Screening of Charges by Multipolar Fluids}
\label{sec:PMF}
In a dipolar fluid, the static dielectric constant is also a measure
of the ability of the fluid to screen charges from one another.  A set
of point charges creates an inhomogeneous field in the fluid, and the
fluid responds to this field as if it was created externally or via
local polarization fluctuations. For this reason, the dielectric
constant can be used to estimate an effective potential between two
point charges ($C_i$ and $C_j$) embedded in the fluid,
\begin{equation}
U_\mathrm{effective} = \frac{C_i C_j}{4 \pi \epsilon_0 \epsilon
  r_{ij}}.
\label{eq:effectivePot}
\end{equation}

The same set of point charges can also create an inhomogeneous field
\textit{gradient}, and this will cause a response in a quadrupolar
fluid that will also cause an effective screening. As discussed above,
however, the relevant physical property in quadrupolar fluids is the
susceptibility, $\chi_Q$.  The screening dielectric associated with
the quadrupolar susceptibility is defined as,\cite{Ernst92}
\begin{equation}
\epsilon = 1 + \chi_Q G = 1 + G \frac{\alpha_Q}{1 + \alpha_Q  B}
\label{eq:dielectricFromQuadrupoles}
\end{equation}
where $G$ is a geometrical factor that depends on the geometry of the
field perturbation,
\begin{equation}
G = \frac{\int_V \left| \nabla \mathbf{E}^\circ \right|^2 d\mathbf{r}}
{\int_V \left|\mathbf{E}^\circ\right|^2 d\mathbf{r}}
\end{equation}
integrated over the interaction volume. Note that this geometrical
factor is also required to compute effective dielectric constants even
when the field gradient is homogeneous over the entire sample.

To measure effective screening in a multipolar fluid, we compute an
effective interaction potential, the potential of mean force (PMF),
between positively and negatively charged ions when they are screened
by the intervening fluid.  The PMF is obtained from a sequence of
simulations in which two ions are constrained to a fixed distance, and
the average constraint force to hold them at a fixed distance $r$ is
collected during a long simulation,\cite{Wilfred07}
\begin{equation}
w(r) = \int_{r_o}^{r}\left\langle \frac{\partial f}{\partial r^\prime}
\right\rangle_{r^\prime} dr^\prime + 2k_BT \ln\left(\frac{r}{r_o}\right) + w(r_o),
\label{eq:pmf}
\end{equation}
where $\braket{\partial f/\partial r^\prime}_{r^\prime}$ is the mean
constraint force required to hold the ions at distance $r^\prime$,
$2k_BT \log(r/r_o)$ is the Fixman factor,\cite{Fixman:1974fk} and
$r_o$ is a reference position (usually taken as a large separation
between the ions). If the dielectric constant is a good measure of the
screening at all inter-ion separations, we would expect $w(r)$ to have
the form in Eq. (\ref{eq:effectivePot}).  Because real fluids are not
continuum dielectrics, the effective dielectric constant is a function
of the interionic separation,
\begin{equation}
\epsilon(r) = \frac{u_\mathrm{raw}(r) - u_\mathrm{raw}(r_o) }{w(r) - w(r_o)} 
\end{equation}
where $u_\mathrm{raw}(r)$ is the direct charge-charge interaction
potential that is in use during the simulation.  $\epsilon(r)$ may
vary considerably from the bulk estimates at short distances, although
it should converge to the bulk value as the separation between the
ions increases.

\section{Simulation Methodology}

To test the formalism developed in the preceding sections, we have
carried out computer simulations using three different techniques: i)
simulations in the presence of external fields, ii) equilibrium
calculations of box moment fluctuations, and iii) potentials of mean
force (PMF) between embedded ions. In all cases, the fluids were
composed of point multipoles protected by a Lennard-Jones potential.
The parameters used in the test systems are given in table
\ref{Tab:C}.

\begin{sidewaystable}
  \caption{\label{Tab:C}The parameters used in simulations to evaluate
    the dielectric response of the new real-space methods.} 
\begin{tabularx}{\textwidth}{r|cc|YYccc|Yccc} \hline
             & \multicolumn{2}{c|}{LJ parameters} &
             \multicolumn{5}{c|}{Electrostatic moments} & & & & \\
 Test system & $\sigma$& $\epsilon$ & $C$ & $D$  &
 $Q_{xx}$ & $Q_{yy}$ & $Q_{zz}$ & mass  & $I_{xx}$ & $I_{yy}$ &
 $I_{zz}$ \\ \cline{6-8}\cline{10-12}
 & (\AA) & (kcal/mol) & (e) & (debye) & \multicolumn{3}{c|}{(debye \AA)} & (amu) & \multicolumn{3}{c}{(amu
 \AA\textsuperscript{2})} \\ \hline
    Dipolar fluid & 3.41 & 0.2381 & - & 1.4026 &-&-&-& 39.948 & 11.613 & 11.613 & 0.0 \\
Quadrupolar fluid & 2.985 & 0.265 & - & - & 0.0 & 0.0 &-2.139 & 18.0153 & 43.0565 & 43.0565 & 0.0  \\
              \ce{q+} & 1.0 & 0.1 & +1 & - & - & - & - & 22.98 & - & - & - \\
              \ce{q-} & 1.0 & 0.1 & -1 & - & - & - & - & 22.98 & - & - & - \\ \hline
\end{tabularx}
\end{sidewaystable}

The first of the test systems consists entirely of fluids of point
dipolar or quadrupolar molecules in the presence of constant field or
field gradients.  Since there are no isolated charges within the
system, the divergence of the field will be zero, \textit{i.e.}
$\nabla \cdot \mathbf{E} = 0$. This condition can be satisfied
by using the relatively simple applied potential as described in the
supplemental material.

When a constant electric field or field gradient is applied to the
system, the molecules align along the direction of the applied field,
and polarize to a degree determined both by the strength of the field
and the fluid's polarizability.  We have calculated ensemble averages
of the box dipole and quadrupole moments as a function of the strength
of the applied fields.  If the fields are sufficiently weak, the
response is linear in the field strength, and one can easily compute
the polarizability directly from the simulations.

The second set of test systems consists of equilibrium simulations of
fluids of point dipolar or quadrupolar molecules simulated in the
absence of any external perturbation. The fluctuation of the ensemble
averages of the box multipolar moment was calculated for each of the
multipolar fluids. The box multipolar moments were computed as simple
sums over the instantaneous molecular moments, and fluctuations in
these quantities were obtained from Eqs. (\ref{eq:flucDip}) and
(\ref{eq:flucQuad}). The macroscopic polarizabilities of the system
were derived using Eqs.(\ref{flucDipole}) and (\ref{flucQuad}).

The final system consists of dipolar or quadrupolar fluids with two
oppositely charged ions embedded within the fluid. These ions are
constrained to be at fixed distance throughout a simulation, although
they are allowed to move freely throughout the fluid while satisfying
that constraint. Separate simulations were run at a range of
constraint distances. A dielectric screening factor was computed using
the ratio between the potential between the two ions in the absence of
the fluid medium and the PMF obtained from the simulations.

We carried out these simulations for all three of the new real-space
electrostatic methods (SP, GSF, and TSF) that were developed in the
first paper (Ref. \onlinecite{PaperI}) in the series. The radius of
the cutoff sphere was taken to be 12~\AA. Each of the real space
methods also depends on an adjustable damping parameter $\alpha$ (in
units of $\mathrm{length}^{-1}$).  We have selected ten different
values of damping parameter: 0.0, 0.05, 0.1, 0.15, 0.175, 0.2, 0.225,
0.25, 0.3, and 0.35~\AA$^{-1}$ in our simulations of the dipolar
liquids, while four values were chosen for the quadrupolar fluids:
0.0, 0.1, 0.2, and 0.3~\AA$^{-1}$.

For each of the methods and systems listed above, a reference
simulation was carried out using a multipolar implementation of the
Ewald sum.\cite{Smith82,Smith98} A default tolerance of
$1 \times 10^{-8}$~kcal/mol was used in all Ewald calculations,
resulting in Ewald coefficient 0.3119~\AA$^{-1}$ for a cutoff radius
of 12~\AA.  All of the electrostatics and constraint methods were
implemented in our group's open source molecular simulation program,
OpenMD,\cite{Meineke05,openmd} which was used for all calculations in
this work.

Dipolar systems contained 2048 Lennard-Jones-protected point dipolar
(Stockmayer) molecules with reduced density $\rho^* = 0.822$,
temperature $T^* = 1.15$, moment of inertia $I^* = 0.025$, and dipole
moment $\mu^* = \sqrt{3.0}$.  These systems were equilibrated for
0.5~ns in the canonical (NVT) ensemble.  Data collection was carried
out over a 1~ns simulation in the microcanonical (NVE) ensemble.  Box
dipole moments were sampled every fs.  For simulations with external
perturbations, field strengths ranging from
0 to 10 $10^{-3}$~V/\AA\ with increments of $ 10^{-4}$~V/\AA\
were carried out for each system.  For dipolar systems the interaction
potential between molecules $i$ and $j$,
\begin{equation}
u_{ij}(\mathbf{r}_{ij}, \mathbf{D}_i, \mathbf{D}_j)  = 4 \epsilon \left(
  \left(\frac{\sigma}{r_{ij}} \right)^{12} - \left(\frac{\sigma}{r_{ij}}
      \right)^{6} \right) - \mathbf{D}_i \cdot
    \mathbf{T}(\mathbf{r}_{ij}) \cdot \mathbf{D}_j
\end{equation}
where the dipole interaction tensor, $\mathbf{T}(\mathbf{r})$, is
given in Eq. (\ref{dipole-diopleTensor}).

Quadrupolar systems contained 4000 linear point quadrupoles with a
density $2.338 \mathrm{~g/cm}^3$ at a temperature of 500~K. These
systems were equilibrated for 200~ps in a canonical (NVT) ensemble.
Data collection was carried out over a 500~ps simulation in the
microcanonical (NVE) ensemble. Components of box quadrupole moments
were sampled every 100 fs. For quadrupolar simulations with external
field gradients, field strengths ranging from
$0 - 9 \times 10^{-2}$~V/\AA$^2$ with increments of
$10^{-2}$~V/\AA$^2$ were carried out for each system.  For quadrupolar
systems the interaction potential between molecules $i$ and $j$,
\begin{equation}
u_{ij}(\mathbf{r}_{ij}, \mathsf{Q}_i, \mathsf{Q}_j)  = 4 \epsilon \left(
  \left(\frac{\sigma}{r_{ij}} \right)^{12} - \left(\frac{\sigma}{r_{ij}}
      \right)^{6} \right) + \mathsf{Q}_i \colon
    \mathbf{T}(\mathbf{r}_{ij}) \colon \mathsf{Q}_j
\end{equation}
where the quadrupole interaction tensor is given in
Eq. (\ref{quadRadial}).

To carry out the PMF simulations, two of the multipolar molecules in
the test system were converted into \ce{q+} and \ce{q-} ions and
constrained to remain at a fixed distance for the duration of the
simulation. The constrained distance was then varied from 5--12~\AA.
In the PMF calculations, all simulations were equilibrated for 500 ps
in the NVT ensemble and run for 5 ns in the microcanonical (NVE)
ensemble.  Constraint forces were sampled every 20~fs.
      
\section{Results}
\subsection{Dipolar fluid}
\begin{figure}
\includegraphics[width=5in]{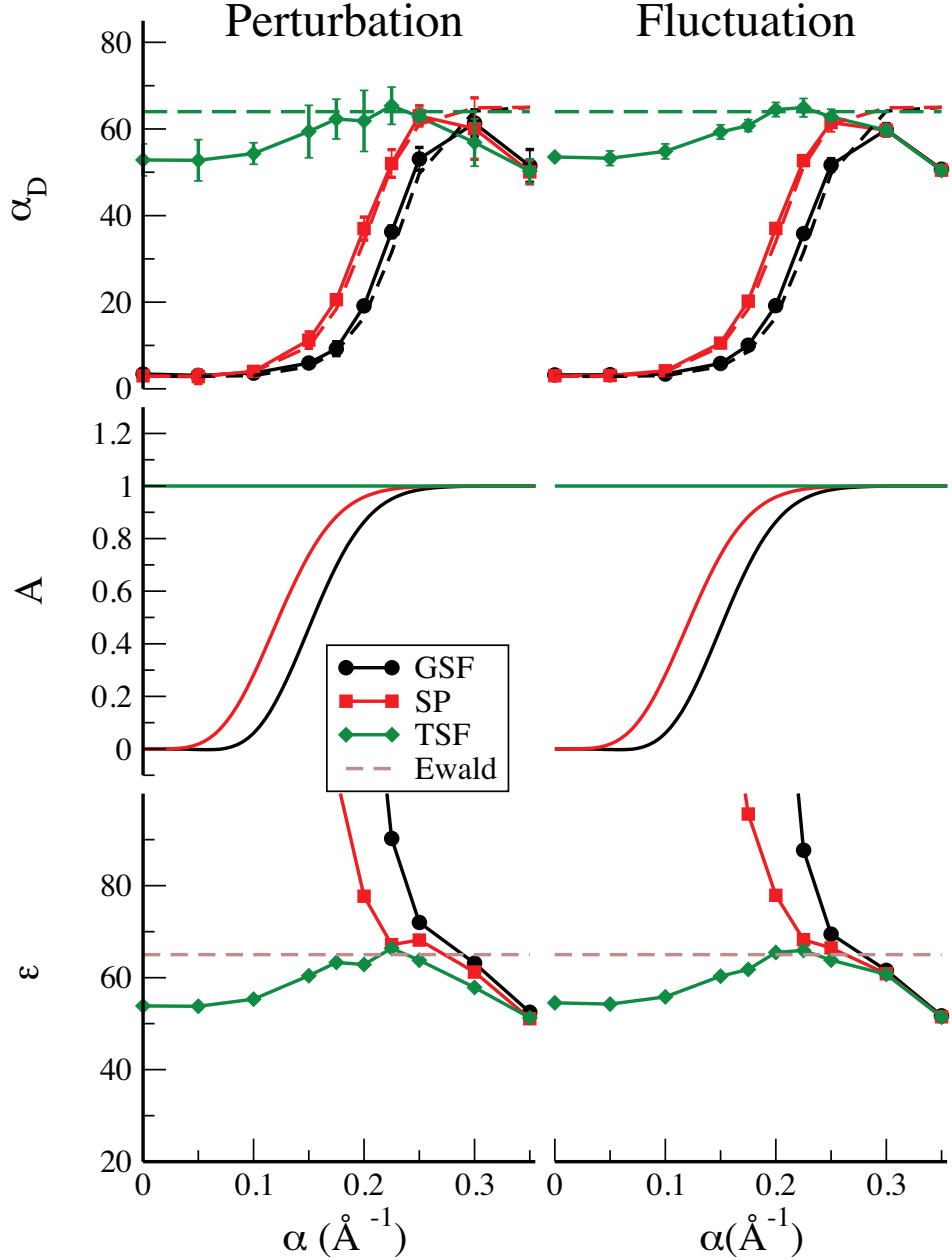}
\caption{The polarizability ($\alpha_D$), correction factor ($A$), and
  dielectric constant ($\epsilon$) for the test dipolar fluid. The
  left panels were computed using external fields, and those on the
  right are the result of equilibrium fluctuations.  In the GSF and SP
  methods, the corrections are large for small values of $\alpha$, and
  an optimal damping coefficient is evident around 0.25 \AA$^{-1}$.
  The dashed lines in the upper panel indicate back-calculation of the
  polarizability using the Ewald estimate (Refs. \onlinecite{Adams81}
  and \onlinecite{NeumannI83}) for the dielectric constant.}
\label{fig:dielectricDipole}
\end{figure}
The bulk polarizability ($\alpha_D$) for the dipolar fluid is
shown in the upper panels in Fig. \ref{fig:dielectricDipole}.  The
polarizability obtained from the both perturbation and fluctuation
approaches are in excellent agreement with each other.  The data also
show a strong dependence on the damping parameter for both the Shifted
Potential (SP) and Gradient Shifted force (GSF) methods, while Taylor
shifted force (TSF) is largely independent of the damping
parameter.

The calculated correction factors discussed in section
\ref{sec:corrFactor} are shown in the middle panels. Because the TSF
method has $A = 1$ for all values of the damping parameter, the
computed polarizabilities need no correction for the dielectric
calculation. The value of $A$ varies with the damping parameter in
both the SP and GSF methods, and inclusion of the correction yields
dielectric estimates (shown in the lower panel) that are generally too
large until the damping reaches $\sim$~0.25~\AA$^{-1}$. Above this
value, the dielectric constants are in reasonable agreement with
previous simulation results.\cite{NeumannI83}

Figure \ref{fig:dielectricDipole} also contains back-calculations of
the polarizability using the reference (Ewald) simulation
results.\cite{NeumannI83} These are indicated with dashed lines in the
upper panels.  It is clear that the expected polarizability for the SP
and GSF methods are quite close to results obtained from the
simulations.  This indicates that the correction formula for the
dipolar fluid (Eq. \ref{correctionFormula}) is extraordinarily
sensitive when the value of $A$ deviates significantly from unity.  It
is also apparent that Gaussian damping is essential for capturing the
field effects from other dipoles. Eq. (\ref{correctionFormula}) works
well when real-space methods employ moderate damping, but is not
capable of providing adequate correction for undamped or weakly-damped
multipoles.

Because the dielectric correction in Eq. (\ref{correctionFormula}) is
so sensitive to $A$ values away from unity, the entries in table
\ref{tab:A} can provide an effective minimum on the values of $\alpha$
that should be used. With a minimum $A=0.995$ and a cutoff radius of
12~\AA, the minimum $\alpha$ values are 0.241~\AA$^{-1}$ (SP) or
0.268~\AA$^{-1}$ (GSF). The TSF method is not sensitive to the choice
of damping parameter. 

\begin{figure}
\includegraphics[width=4in]{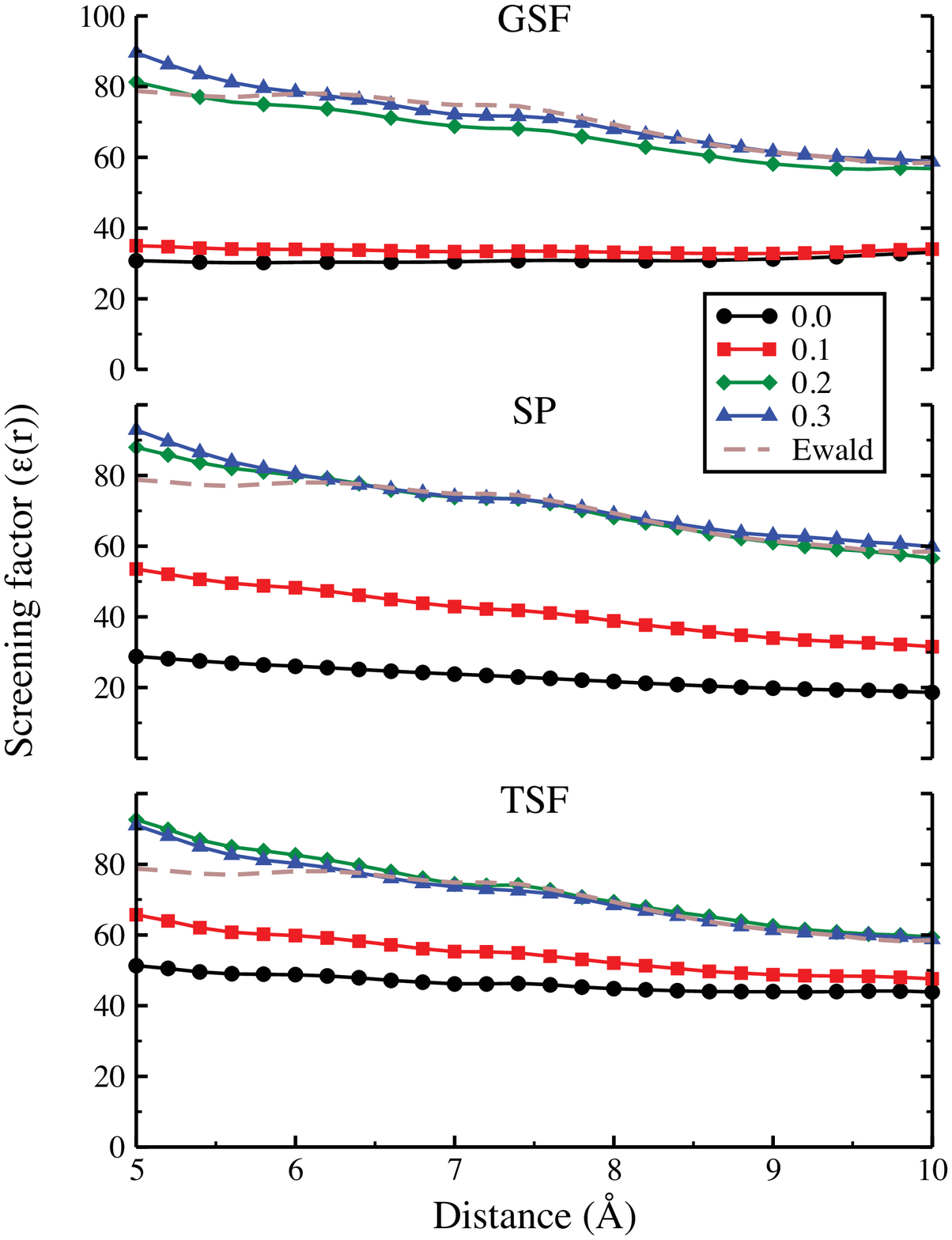}
\caption{The distance-dependent screening factor, $\epsilon(r)$,
  between two ions immersed in the dipolar fluid. The new methods are
  shown in separate panels, and different values of the damping
  parameter ($\alpha$) are indicated with different symbols. All of
  the methods appear to be converging to the bulk dielectric constant
  ($\sim 65$) for higher values of $\alpha$ and at large ion
  separations.}
\label{fig:ScreeningFactor_Dipole}
\end{figure}
We have also evaluated the distance-dependent screening factor,
$\epsilon(r)$, between two oppositely charged ions when they are
placed in the dipolar fluid.  These results were computed using
Eq. \ref{eq:pmf} and are shown in Fig.
\ref{fig:ScreeningFactor_Dipole}.

The screening factor is similar to the dielectric constant, but
measures a local property of the ions in the fluid and depends on both
ion-dipole and dipole-dipole interactions. These interactions depend
on the distance between ions as well as the electrostatic interaction
methods utilized in the simulations. The screening should converge to
the dielectric constant when the field due to ions is small. This
occurs when the ions are separated (or when the damping parameter is
large). In Fig. \ref{fig:ScreeningFactor_Dipole} we observe that for
the higher value of damping alpha \textit{i.e.}
$\alpha > 0.2$~\AA$^{-1}$ and large separation between ions, the
screening factor does indeed approach the correct dielectric constant.

It is also notable that the TSF method again displays smaller
perturbations away from the correct dielectric screening behavior.  We
also observe that for TSF, the method yields high dielectric screening
even for lower values of $\alpha$.

At short distances, the presence of the ions creates a strong local
field that acts to align nearby dipoles nearly perfectly in opposition
to the field from the ions.  This has the effect of increasing the
effective screening when the ions are brought close to one another.
This effect is present even in the full Ewald treatment, and indicates
that the local ordering behavior is being captured by all of the
moderately-damped real-space methods.

\subsubsection*{Distance-dependent Kirkwood factors}
\begin{figure}
\includegraphics[width=\linewidth]{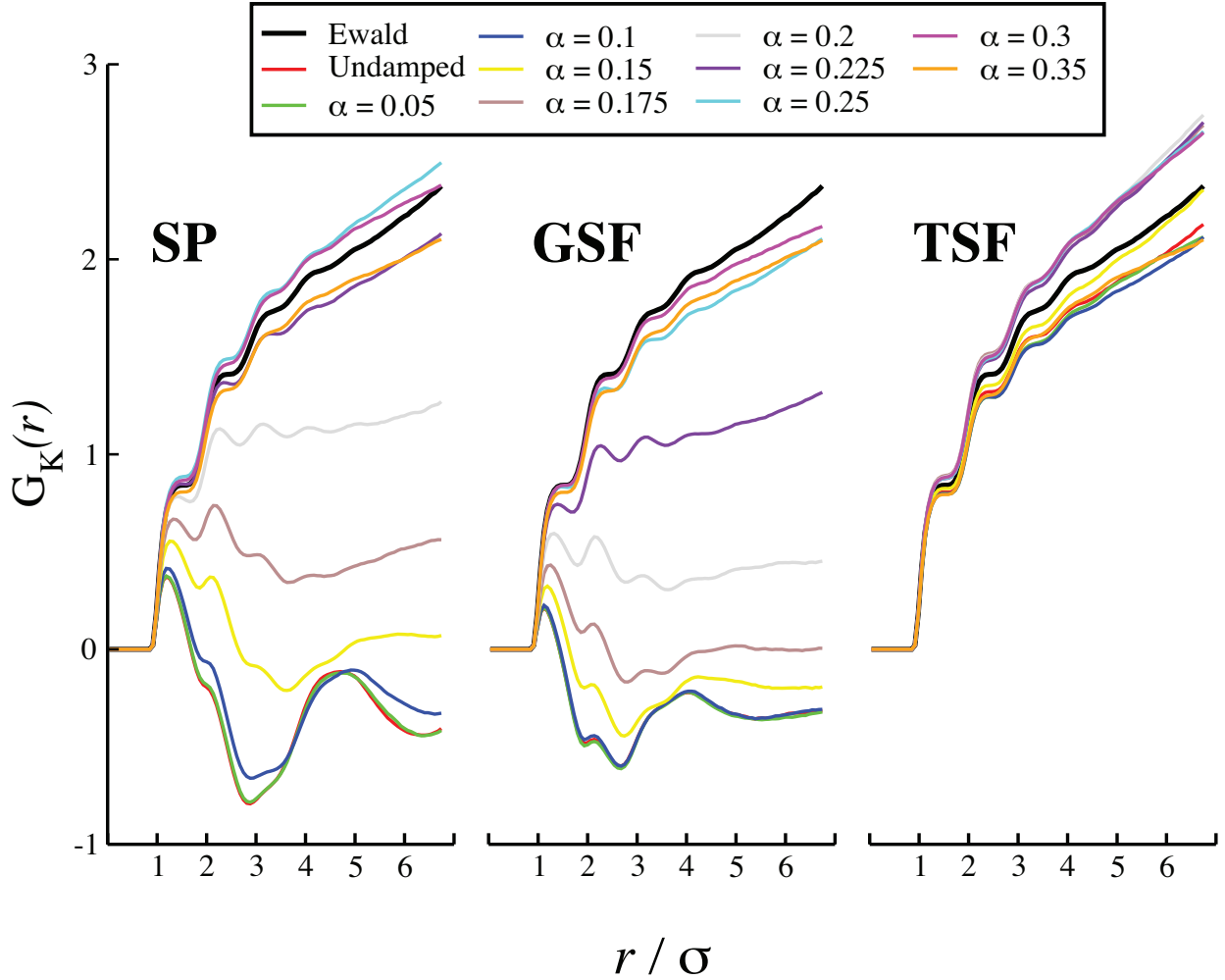}
\caption{The distance-dependent Kirkwood factors of the dipolar system
  for the three real space methods at a range of Gaussian damping
  parameters ($\alpha$) with a cutoff $r_c = 3.52 \sigma$.}
\label{fig:kirkwood}
\end{figure}

One of the most sensitive measures of dipolar ordering in a liquid is
the disance dependent Kirkwood factor,
\begin{equation}
G_K(r) = \left< \frac{1}{N} \sum_{i} \sum_{\substack{j \\ r_{ij} < r}}
  \frac{\mathbf{D}_i \cdot \mathbf{D}_j}{\left| D_i \right| \left| D_j
    \right|} \right> 
\label{eq:kirkwood}
\end{equation}
which measures the net orientational (cosine) ordering of dipoles
inside a sphere of radius $r$.  The outer brackets denote a
configurational average.  Figure \ref{fig:kirkwood} shows $G_K(r)$ for
the three real space methods with $r_c = 3.52 \sigma = 12$~\AA~ and
for the Ewald sum.  These results were obtained from unperturbed 5~ns
simulations of the dipolar fluid in the microcanonical (NVE)
ensemble. For SP and GSF, the underdamped cases exhibit the ``hole''
at $r_c$ that is sometimes seen in cutoff-based method simulations of
liquid water,\cite{Mark:2002rt,Fukuda:2012fr} but for values of
$\alpha > 0.225$ \AA$^{-1}$, agreement with the Ewald results is good.
Note that like the dielectric constant, $G_K(r)$ can also be corrected
using the expressions for $A$ in table \ref{tab:A}. This is discussed
in more detail in the supplemental material.

\subsection{Quadrupolar fluid}
\begin{figure}
\includegraphics[width=4in]{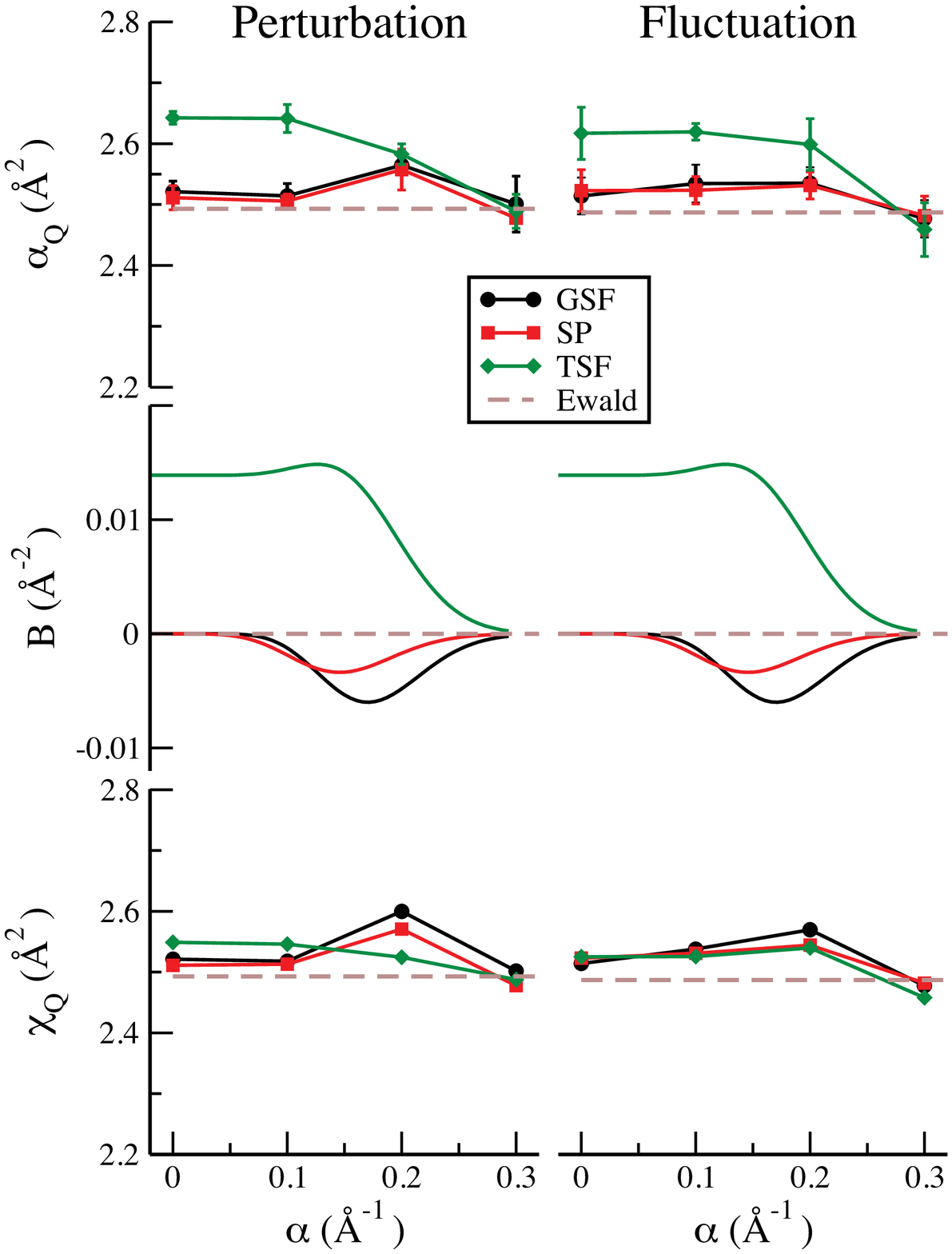}
\caption{The quadrupole polarizability ($\alpha_Q$), correction factor
  ($B$), and susceptibility ($\chi_Q$) for the test quadrupolar
  fluid. The left panels were computed using external field gradients,
  and those on the right are the result of equilibrium fluctuations.
  The GSF and SP methods allow nearly unmodified use of the
  ``conducting boundary'' or polarizability results in place of the
  bulk susceptibility.}
\label{fig:dielectricQuad}
\end{figure}
The polarizability ($\alpha_Q$), correction factor ($\mathrm{B}$), and
susceptibility ($\chi_Q$) for the quadrupolar fluid is plotted against
damping parameter Fig.  \ref{fig:dielectricQuad}.  In quadrupolar
fluids, both the polarizability and susceptibility have units of
$\mathrm{length}^2$. Although the susceptibility has dimensionality,
it is the relevant measure of macroscopic quadrupolar
properties.\cite{JeonI03, JeonII03} The left panel in
Fig. \ref{fig:dielectricQuad} shows results obtained from the applied
field gradient simulations whereas the results from the equilibrium
fluctuation formula are plotted in the right panels.

The susceptibility for the quadrupolar fluid is obtained from
quadrupolarizability and a correction factor using
Eq. (\ref{eq:quadrupolarSusceptiblity}).  The susceptibilities are
shown in the bottom panels of Fig. \ref{fig:dielectricQuad}. All three
methods: (SP, GSF, and TSF) produce small correction factors,
$\mathrm{B}$, so all show similar susceptibilities over the range of
damping parameters.  This shows that susceptibility derived using the
quadrupolarizability and the correction factors are essentially
independent of the electrostatic method utilized in the simulation.

There is a notable difference in the dependence on $\alpha$ for the
quadrupolar correction compared with the dipolar correction. This is
due to the reduced range of the quadrupole-quadrupole interaction when
compared with dipolar interactions. The effects of the Gaussian
damping for dipoles are significant near the cutoff radius, which can
be observed in Fig. \ref{fig:kirkwood}, while for quadrupoles, most of
the interaction is naturally diminished by that point. Because
overdamping can obscure orientational preferences, quadrupolar fluids
can be safely simulated with smaller values of $\alpha$ than a similar
dipolar fluid.

\begin{figure}
\includegraphics[width=\linewidth]{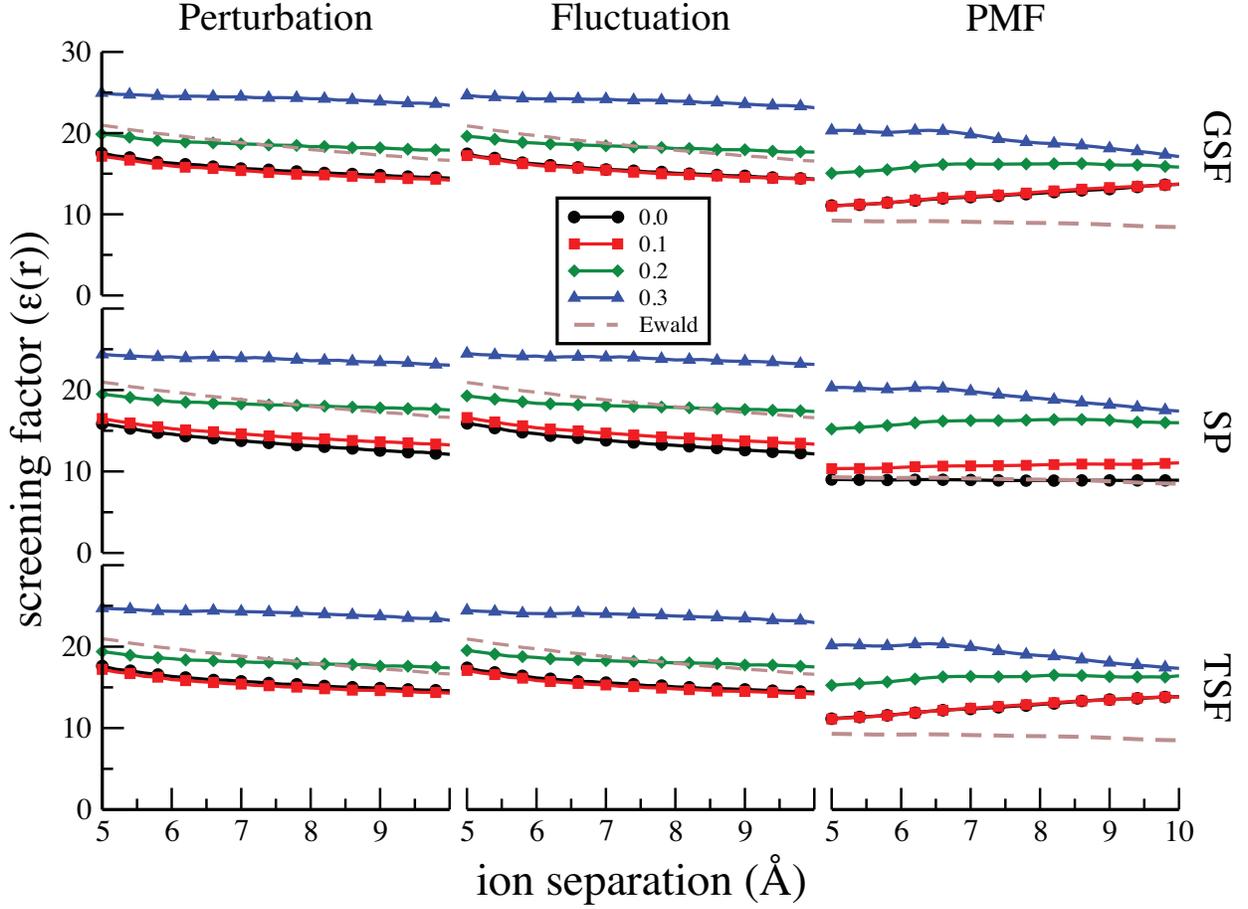}
\caption{\label{fig:screenQuad} The distance-dependent screening
  factor, $\epsilon(r)$, between two ions immersed in the quadrupolar
  fluid. Results from the perturbation and fluctuation methods are
  shown in left and central panels. Here the susceptibility is
  calculated from the bulk simulations and the geometrical factor is
  evaluated using Eq. (\ref{eq:geometricalFactor}) using the field and
  field-gradient produced by the two ions. The right hand panel shows
  the screening factor obtained from the PMF calculations.}
\end{figure}
A more difficult test of the quadrupolar susceptibility is made by
comparing with direct calculation of the electrostatic screening using
the potential of mean force (PMF). Since the effective dielectric
constant for a quadrupolar fluid depends on the geometry of the field
and field gradient, this is not a physical property of the quadrupolar
fluid. 

The geometrical factor for embedded ions changes with the ion
separation distance. It is therefore reasonable to treat the
dielectric constant as a distance-dependent screening factor.  Since
the quadrupolar molecules couple with the gradient of the field, the
distribution of the quadrupoles will be inhomogeneously distributed
around the point charges.  Hence the distribution of quadrupolar
molecules should be taken into account when computing the geometrical
factors in the presence of this perturbation,
\begin{eqnarray}
G &=& \frac{\int_V g(\mathbf{r})  \left|\nabla \mathbf{E}^\circ
  \right|^2 d\mathbf{r}}{\int_V\left|\mathbf{E}^\circ\right|^2
  d\mathbf{r}}  \nonumber \\ \nonumber \\
 &=& \frac{ 2 \pi \int_{-1}^{1}  \int_{0}^{R} r^2  g(r,
     \cos\theta)  \left|\nabla \mathbf{E}^\circ  \right|^2 dr d(\cos\theta) }{\int_V\left|\mathbf{E}^\circ\right|^2 d\mathbf{r}}
\label{eq:geometricalFactor}
\end{eqnarray}
where $g(r,\cos\theta)$ is a distribution function for the quadrupoles
with respect to an origin at midpoint of a line joining the two probe
charges.

The effective screening factor is plotted against ion separation
distance in Fig. \ref{fig:screenQuad}.  The screening evaluated from
the perturbation and fluctuation methods are shown in the left and
central panels. Here the susceptibilities are calculated from bulk
fluid simulations and the geometrical factors are evaluated using the
field and field gradients produced by the ions. The field gradients
have been weighted by the $g(r, \cos\theta)$ from the PMF calculations
(Eq. (\ref{eq:geometricalFactor})).  The right hand panel shows the
screening factor obtained directly from the PMF calculations.

We note that the screening factor obtained from both the perturbation
and fluctuation methods are in good agreement with each other at
similar values of $\alpha$, and agree with Ewald for
$\alpha = 0.2$~\AA$^{-1}$.  The magnitude of these screening factors
depends strongly on the $g(r, \cos\theta)$ weighting originating in
the PMF calculations.

In Ewald-based simulations, the PMF calculations include interactions
between periodic replicas of the ions, and there is a significant
reduction in the screening factor because of this effect.  Because the
real-space methods do not include coupling to periodic replicas, both
the magnitude and distance-dependent decay of the PMF are
significantly larger.  For moderate damping ($\alpha \sim 0.2-0.3$
\AA$^{-1}$), screening factors for GSF, TSF, and SP are converging to
similar values at large ion separations, and this value is the same as
the large-separation estimate from the perturbation and fluctuation
simulations for $\alpha \sim 0.2$ \AA$^{-1}$.  The PMF calculations
also show signs of coalescence of the ion solvation shells at
separations smaller than 7~\AA.  At larger separations, the
$\alpha = 0.2$~\AA$^{-1}$ PMF calculations appear to be reproducing
the bulk screening values.  These results suggest that using either
TSF or GSF with moderate damping is a relatively safe way to predict
screening in quadrupolar fluids.

\section{Conclusions}
We have used both perturbation and fluctuation approaches to evaluate
dielectric properties for dipolar and quadrupolar fluids.  The static
dielectric constant is the relevant bulk property for dipolar fluids,
while the quadrupolar susceptibility plays a similar role for
quadrupoles.  Corrections to both the static dielectric constant and
the quadrupolar susceptibility were derived for three new real space
electrostatic methods, and these corrections were tested against a
third measure of dielectric screening, the potential of mean force
between two ions immersed in the fluids.

For the dipolar fluids, we find that the polarizability evaluated
using the perturbation and fluctuation methods show excellent
agreement, indicating that equilibrium calculations of the dipole
fluctuations are good measures of bulk polarizability. 

One of the findings of the second paper in this series is that the
moderately damped GSF and SP methods were most suitable for molecular
dynamics and Monte Carlo simulations, respectively.\cite{PaperII} Our
current results show that dielectic properties like $\epsilon$ and
$G_K(r)$ are sensitive probes of local treatment of electrostatic
damping for the new real space methods, as well as for the Ewald
sum. Choosing a Gaussian damping parameter ($\alpha$) in a reasonable
range is therefore essential for obtaining agreement between the
electrostatic methods. A physical explanation of this rests on the
local orientational preferences of other molecules around a central
dipole.  The orientational contributions to dipolar interactions are
weighted by two radial functions ($v_{21}(r)$ and $v_{22}(r)$).  The
relative magnitudes of these functions, and therefore the
orientational preferences of local dipoles, are quite sensitive to the
value of $\alpha$.  With moderate damping, the ratio approaches the
orientational preferences of Ewald-based simulations, removing the
``hole'' in $G_K(r)$ for underdamped SP and GSF simulations
(Fig. \ref{fig:kirkwood}).

The derived correction formulae can approximate bulk properties from
non-optimal parameter choices, as long as the methods are used in a
relatively ``safe'' range of damping. The newly-derived entries in
table \ref{tab:A} can provide an effective minimum on the values of
$\alpha$ that should be used in simulations.  With a cutoff radius of
12 \AA, $\alpha = 0.241$ \AA$^{-1}$ (SP) or $0.268$ \AA$^{-1}$ (GSF)
would capture dielectric screening with reasonable fidelity. The
sensitivity of the dielectric screening is also observed in the
effective screening of ions embedded in the fluid.

With good choices of $\alpha$, the dielectric constant evaluated using
the computed polarizability and correction factors agrees well with
the previous Ewald-based simulation results.\cite{Adams81,NeumannI83}
Although the TSF method alters many dynamic and structural features in
multipolar liquids,\cite{PaperII} it is surprisingly good at computing
bulk dielectric properties at nearly all ranges of the damping
parameter.  In fact, the correction factor, $A = 1$, for the TSF
method so the conducting boundary formula is essentially correct when
using this method for point dipolar fluids.

As in the dipolar case, the quadpole polarizability evaluated from
both perturbation and fluctuation simulations show good agreement,
again confirming that equilibrium fluctuation calculations are
sufficient to reproduce bulk dielectric properties in these fluids.
The quadrupolar susceptibility calculated via our derived correction
factors produces similar results for all three real space
methods. Similarly, with good choices of the damping parameter, the
screening factor calculated using the susceptibility and a weighted
geometric factor provides good agreement with results obtained
directly via potentials of mean force.  For quadrupolar fluids, the
distance dependence of the electrostatic interaction is significantly
reduced and the correction factors are all small.  These points
suggest that how an electrostatic method treats the cutoff radius
become less consequential for higher order multipoles.
   
For this reason, our recommendation is that the moderately-damped
($\alpha = 0.25-0.27$~\AA$^{-1}$) GSF method is a good choice for
molecular dynamics simulations where point-multipole interactions are
being utilized to compute bulk dielectric properties of fluids.  

\section*{Supplementary Material} 
See supplementary material for information on interactions with
spatially varying fields, Boltzmann averages, self-contributions from
quadrupoles, and corrections to distance-dependent Kirkwood factors.

\begin{acknowledgments}
  Support for this project was provided by the National Science Foundation
  under grant CHE-1362211. Computational time was provided by the
  Center for Research Computing (CRC) at the University of Notre
  Dame. The authors would like to thank the reviewer for helpful
  comments and suggestions.
\end{acknowledgments}

\appendix
\section{Contraction of the quadrupolar tensor with the traceless
  quadrupole moment }
\label{ap:quadContraction}
For quadrupolar liquids modeled using point quadrupoles, the
interaction tensor is shown in Eq. (\ref{quadRadial}).  The Fourier
transformation of this tensor for $ \mathbf{k} = 0$ is,
\begin{equation}
\tilde{T}_{\alpha\beta\gamma\delta}(0) = \int_V T_{\alpha\beta\gamma\delta}(\mathbf{r}) d \mathbf{r}
\end{equation}
On the basis of symmetry, the 81 elements can be placed in four
different groups: $\tilde{T}_{aaaa}$, $\tilde{T}_{aaab}$,
$\tilde{T}_{aabb}$, and $\tilde{T}_{aabc}$, where $a$, $b$, and $c$,
and can take on distinct values from the set $\left\{x, y, z\right\}$.
The elements belonging to each of these groups can be obtained using
permutations of the indices.  Integration of all of the elements shows
that only the groups with indices ${aaaa}$ and ${aabb}$ are non-zero.

We can derive values of the components of $\tilde{T}_{aaaa}$ and
$\tilde{T}_{aabb}$ as follows;
\begin{eqnarray}
\tilde{T}_{xxxx}(0) &=&
\int_{\textrm{V}} 
\left[ 3v_{41}(r)+6x^2v_{42}(r)/r^2 + x^4\,v_{43}(r)/r^4 \right] d\mathbf{r} \nonumber \\ 
&=&12\pi \int_0^{r_c} 
\left[ v_{41}(r)+\frac{2}{3} v_{42}(r) + \frac{1}{15}v_{43}(r) \right] r^2\,dr =
\mathrm{12 \pi B}
\end{eqnarray}
and 
\begin{eqnarray}
  \tilde{T}_{xxyy}(0)&=&
                         \int_{\textrm{V}} 
                         \left[ v_{41}(r)+(x^2+y^2) v_{42}(r)/r^2 + x^2 y^2\,v_{43}(r)/r^4 \right] d\mathbf{r} \nonumber \\
                     &=&4\pi \int_0^{r_c}
                         \left[ v_{41}(r)+\frac{2}{3} v_{42}(r) + \frac{1}{15}v_{43}(r) \right] r^2\,dr =
                         \mathrm{4 \pi B}.
\end{eqnarray}
These integrals yield the same values for all permutations of the
indices in both tensor element groups.  In Eq.
\ref{fourierQuadZeroK}, for a particular value of the quadrupolar
polarization $\tilde{\Theta}_{aa}$ we can contract
$\tilde{T}_{aa\gamma\delta}(0)$ with $\tilde{\Theta}_{\gamma\delta}$,
using the traceless properties of the quadrupolar moment,
\begin{eqnarray}
\tilde{T}_{xx\gamma\delta}(0)\tilde{\Theta}_{\gamma\delta}(0) &=& \tilde{T}_{xxxx}(0)\tilde{\Theta}_{xx}(0) + \tilde{T}_{xxyy}(0)\tilde{\Theta}_{yy}(0) + \tilde{T}_{xxzz}(0)\tilde{\Theta}_{zz}(0) \nonumber \\
&=& 12 \pi \mathrm{B}\tilde{\Theta}_{xx}(0) +
    4 \pi \mathrm{B}\tilde{\Theta}_{yy}(0) +
    4 \pi \mathrm{B}\tilde{\Theta}_{zz}(0) \nonumber \\
&=& 8 \pi \mathrm{B}\tilde{\Theta}_{xx}(0) + 4 \pi
    \mathrm{B}\left(\tilde{\Theta}_{xx}(0)+\tilde{\Theta}_{yy}(0) +
    \tilde{\Theta}_{zz}(0)\right) \nonumber \\
&=& 8 \pi \mathrm{B}\tilde{\Theta}_{xx}(0)
\end{eqnarray} 
Similarly for a quadrupolar polarization $\tilde{\Theta}_{xy}$ in
Eq. \ref{fourierQuadZeroK}, we can contract
$\tilde{T}_{xy\gamma\delta}(0)$ with $\tilde{\Theta}_{\gamma\delta}$,
using the only surviving terms of the tensor, 
\begin{eqnarray}
\tilde{T}_{xy\gamma\delta}(0)\tilde{\Theta}_{\gamma\delta}(0) &=& \tilde{T}_{xyxy}(0)\tilde{\Theta}_{xy}(0) + \tilde{T}_{xyyx}(0)\tilde{\Theta}_{yx}(0) \nonumber \\
&=& 4 \pi \mathrm{B}\tilde{\Theta}_{xy}(0) +
    4 \pi \mathrm{B}\tilde{\Theta}_{yx}(0) \nonumber \\
&=& 8 \pi \mathrm{B}\tilde{\Theta}_{xy}(0)
\end{eqnarray}
Here, we have used the symmetry of the quadrupole tensor to combine
the symmetric terms. Therefore we can write matrix contraction for
$\tilde{T}_{\alpha\beta\gamma\delta}(\mathrm{0})$ and
$ \tilde{\Theta}_{\gamma\delta}(\mathrm{0})$ in a general form,
\begin{equation}
\tilde{T}_{\alpha\beta\gamma\delta}(\mathrm{0})\tilde{\Theta}_{\gamma\delta}(\mathrm{0})
= 8 \pi \mathrm{B} \tilde{\Theta}_{\alpha\beta}(\mathrm{0}),
\label{contract}
\end{equation}
which is the same as Eq. (\ref{quadContraction}).

When the molecular quadrupoles are represented by point charges, the
symmetry of the quadrupolar tensor is same as for point quadrupoles
(see Eqs.~\ref{quadCharge} and \ref{quadRadial}). However, for
molecular quadrupoles represented by point dipoles, the symmetry of
the quadrupolar tensor must be handled separately (compare
Eqs.~\ref{quadDip} and~\ref{quadRadial}). Although there is a
difference in symmetry, the final result (Eq.~\ref{contract}) also holds
true for dipolar representations.

\section{Quadrupolar correction factor for the Ewald-Kornfeld (EK)
  method}
The interaction tensor between two point quadrupoles in the Ewald
method may be expressed,\cite{Smith98,NeumannII83}
\begin{align}
{T}_{\alpha\beta\gamma\delta}(\mathbf{r}) = &\frac{4\pi}{V
                                              }\sum_{k\neq0}^{\infty}
                                              e^{-k^2 / 4
                                              \kappa^2} e^{-i\mathbf{k}\cdot
                                              \mathbf{r}} \left(\frac{r_\alpha r_\beta k_\delta k_\gamma}{k^2}\right)  \nonumber \\
&+ \left(\delta_{\alpha\beta}\delta_{\gamma\delta}+\delta_{\alpha\gamma}\delta_{\beta\delta}+\delta_{\alpha\delta}\delta_{\beta\gamma}\right) 
B_2(r) \nonumber \\
&- \left(\delta_{\gamma\delta} r_\alpha r_\beta +  \mathrm{ 5\; permutations}\right) B_3(r) \nonumber \\
&+ \left(r_\alpha r_\beta r_\gamma r_\delta \right)  B_4(r)
\label{ewaldTensor}
\end{align}
where $B_n(r)$ are radial functions defined in reference
\onlinecite{Smith98},
\begin{align}
B_2(r)  =& \frac{3}{r^5} \left(\frac{2r\kappa e^{-r^2 \kappa^2}}{\sqrt{\pi}}+\frac{4r^3\kappa^3 e^{-r^2 \kappa^2}}{3\sqrt{\pi}}+\mathrm{erfc(\kappa r)} \right) \\
B_3(r) =& - \frac{15}{r^7}\left(\frac{2r\kappa e^{-r^2 \kappa^2}}{\sqrt{\pi}}+\frac{4r^3\kappa^3 e^{-r^2 \kappa^2}}{3\sqrt{\pi}}+\frac{8r^5\kappa^5 e^{-r^2 \kappa^2}}{15\sqrt{\pi}}+\mathrm{erfc(\kappa r)} \right) \\
B_4(r) =& \frac{105}{r^9}\left(\frac{2r\kappa e^{-r^2
          \kappa^2}}{\sqrt{\pi}}+\frac{4r^3\kappa^3 e^{-r^2 \kappa^2}}{3\sqrt{\pi}}+\frac{8r^5\kappa^5 e^{-r^2 \kappa^2}}{15\sqrt{\pi}}
 + \frac{16r^7\kappa^7 e^{-r^2 \kappa^2}}{105\sqrt{\pi}} +  \mathrm{erfc(\kappa r)} \right)
\end{align}  

We can divide ${T}_{\alpha\beta\gamma\delta}(\mathbf{r})$ into three
parts:
\begin{eqnarray}
& & \mathbf{T}(\mathbf{r}) =
    \mathbf{T}^\mathrm{K}(\mathbf{r}) +
    \mathbf{T}^\mathrm{R1}(\mathbf{r}) +
    \mathbf{T}^\mathrm{R2}(\mathbf{r}) 
\end{eqnarray}
where the first term is the reciprocal space portion.  Since the
quadrupolar correction factor $B = \tilde{T}_{abab}(0) / 4\pi$ and
$\mathbf{k} = 0 $ is excluded from the reciprocal space sum,
$\mathbf{T}^\mathrm{K}$ will not contribute.\cite{NeumannII83} The
remaining terms,
\begin{equation}
\mathbf{T}^\mathrm{R1}(\mathbf{r}) =  \mathbf{T}^\mathrm{bare}(\mathbf{r}) \left(\frac{2r\kappa e^{-r^2
          \kappa^2}}{\sqrt{\pi}}+\frac{4r^3\kappa^3 e^{-r^2 \kappa^2}}{3\sqrt{\pi}}+\frac{8r^5\kappa^5 e^{-r^2 \kappa^2}}{15\sqrt{\pi}}
 + \frac{16r^7\kappa^7 e^{-r^2 \kappa^2}}{105\sqrt{\pi}} +  \mathrm{erfc(\kappa r)} \right)
\end{equation}
and
\begin{eqnarray}
T^\mathrm{R2}_{\alpha\beta\gamma\delta}(\mathbf{r}) =  &+& \left(\delta_{\gamma\delta} r_\alpha r_\beta +  \mathrm{ 5\; permutations}\right) \frac{16 \kappa^7 e^{-r^2 \kappa^2}}{7\sqrt{\pi}} \nonumber \\
 &-&\left(\delta_{\alpha\beta}\delta_{\gamma\delta}+\delta_{\alpha\gamma}\delta_{\beta\delta}+\delta_{\alpha\delta}\delta_{\beta\gamma}\right) \left(\frac{8 \kappa^5 e^{-r^2 \kappa^2}}{5\sqrt{\pi}}+ \frac{16 r^2\kappa^7 e^{-r^2 \kappa^2}}{35\sqrt{\pi} }\right)
\end{eqnarray}
are contributions from the real space
sum.\cite{Adams76,Adams80,Adams81} Here
$\mathbf{T}^\mathrm{bare}(\mathbf{r})$ is the unmodified quadrupolar
tensor (for undamped quadrupoles).  Due to the angular symmetry of the
unmodified tensor, the integral of
$\mathbf{T}^\mathrm{R1}(\mathbf{r})$ will vanish when integrated over
a spherical region. The only term contributing to the correction
factor (B) is therefore
$T^\mathrm{R2}_{\alpha\beta\gamma\delta}(\mathbf{r})$, which allows us
to derive the correction factor for the Ewald-Kornfeld (EK) method,
\begin{eqnarray}
\mathrm{B} &=& \frac{1}{4\pi} \int_V T^\mathrm{R2}_{abab}(\mathbf{r}) \nonumber \\
&=& -\frac{8r_c^3 \kappa^5 e^{-\kappa^2 r_c^2}}{15\sqrt{\pi}}.
\end{eqnarray}

\newpage
%\bibliography{dielectric_new}

\end{document}

\title{Supplemental Material for: Real space electrostatics for
  multipoles. III. Dielectric Properties}

\author{Madan Lamichhane}
\affiliation{Department of Physics, University
of Notre Dame, Notre Dame, IN 46556}
\author{Thomas Parsons}
\affiliation{Department of Chemistry and Biochemistry, University
of Notre Dame, Notre Dame, IN 46556}
\author{Kathie E. Newman}
\affiliation{Department of Physics, University
of Notre Dame, Notre Dame, IN 46556}
\author{J. Daniel Gezelter}
\email{gezelter@nd.edu.}
\affiliation{Department of Chemistry and Biochemistry, University
of Notre Dame, Notre Dame, IN 46556}

\date{\today}% It is always \today, today,

\begin{abstract}
  This document includes useful relationships for computing the
  interactions between fields and field gradients and point multipolar
  representations of molecular electrostatics. We also provide
  explanatory derivations of a number of relationships used in the
  main text. This includes the Boltzmann averages of quadrupole
  orientations, and the interaction of a quadrupole density with the
  self-generated field gradient. This last relationship is assumed to
  be zero in the main text but is explicitly shown to be zero here.  A
  discussion of method-dependent corrections to the distance-dependent
  Kirkwood factors is also included.
\end{abstract}

\maketitle

\section{Generating Uniform Field Gradients}
One important task in carrying out the simulations mentioned in the
main text was to generate uniform electric field gradients.  To do
this, we relied heavily on both the notation and results from Torres
del Castillo and Mend\'{e}z Garido.\cite{Torres-del-Castillo:2006uo}
In this work, tensors were expressed in Cartesian components, using at
times a dyadic notation. This proves quite useful for computer
simulations that make use of toroidal boundary conditions.

An alternative formalism uses the theory of angular momentum and
spherical harmonics and is common in standard physics texts such as
Jackson,\cite{Jackson98} Morse and Feshbach,\cite{Morse:1946zr} and
Stone.\cite{Stone:1997ly} Because this approach has its own
advantages, relationships are provided below comparing that
terminology to the Cartesian tensor notation.

The gradient of the electric field,
\begin{equation*}
\mathsf{G}(\mathbf{r}) = -\nabla \nabla \Phi(\mathbf{r}),
\end{equation*}
where $\Phi(\mathbf{r})$ is the electrostatic potential.  In a
charge-free region of space, $\nabla \cdot \mathbf{E}=0$, and
$\mathsf{G}$ is a symmetric traceless tensor.  From symmetry
arguments, we know that this tensor can be written in terms of just
five independent components.

Following Torres del Castillo and Mend\'{e}z Garido's notation, the
gradient of the electric field may also be written in terms of two
vectors $\mathbf{a}$ and $\mathbf{b}$,
\begin{equation*}
G_{ij}=\frac{1}{2} (a_i b_j + a_j b_i) - \frac{1}{3}(\mathbf a \cdot \mathbf b) \delta_{ij} .
\end{equation*} 
If the vectors $\mathbf{a}$ and $\mathbf{b}$ are unit vectors, the
electrostatic potential that generates a uniform gradient may be
written:
\begin{align}
\Phi(x, y, z) =\; -\frac{g_o}{2} & \left(\left(a_1b_1 -
                         \frac{cos\psi}{3}\right)\;x^2+\left(a_2b_2
                         - \frac{cos\psi}{3}\right)\;y^2 +
                         \left(a_3b_3 -
                         \frac{cos\psi}{3}\right)\;z^2 \right. \nonumber \\
 & + (a_1b_2 + a_2b_1)\; xy + (a_1b_3 + a_3b_1)\; xz + (a_2b_3 + a_3b_2)\; yz \bigg) .
\label{eq:appliedPotential}
\end{align} 
Note $\mathbf{a}\cdot\mathbf{a} = \mathbf{b} \cdot \mathbf{b} = 1$,
$\mathbf{a} \cdot \mathbf{b}=\cos \psi$, and $g_0$ is the overall
strength of the potential.

Taking the gradient of Eq. (\ref{eq:appliedPotential}), we find the
field due to this potential,
\begin{equation}
\mathbf{E} = -\nabla \Phi
=\frac{g_o}{2} \left(\begin{array}{ccc}
2(a_1 b_1 - \frac{cos\psi}{3})\; x & +\; (a_1 b_2 + a_2 b_1)\; y & +\; (a_1 b_3 + a_3 b_1)\; z \\
 (a_2 b_1 + a_1 b_2)\; x & +\; 2(a_2 b_2 - \frac{cos\psi}{3})\; y & +\;  (a_2 b_3 + a_3 b_3)\; z \\
(a_3 b_1 + a_3 b_2)\; x & +\;  (a_3 b_2 + a_2 b_3)\; y & +\; 2(a_3 b_3 - \frac{cos\psi}{3})\; z 
\end{array} \right),
\label{eq:CE}
\end{equation} 
while the gradient of the electric field in this form,
\begin{equation}
\mathsf{G} = \nabla\mathbf{E} 
= \frac{g_o}{2}\left(\begin{array}{ccc}
2(a_1\; b_1 - \frac{cos\psi}{3}) &  (a_1\; b_2 \;+ a_2\; b_1) & (a_1\; b_3 \;+ a_3\; b_1) \\
 (a_2\; b_1 \;+ a_1\; b_2) & 2(a_2\; b_2 \;- \frac{cos\psi}{3}) & (a_2\; b_3 \;+ a_3\; b_3) \\
(a_3\; b_1 \;+ a_3\; b_2) & (a_3\; b_2 \;+ a_2\; b_3) & 2(a_3\; b_3 \;- \frac{cos\psi}{3})
\end{array} \right),
\label{eq:GC}
\end{equation}  
is uniform over the entire space.  Therefore, to describe a uniform
gradient in this notation, two unit vectors ($\mathbf{a}$ and
$\mathbf{b}$) as well as a potential strength, $g_0$, must be
specified. As expected, this requires five independent parameters.

The common alternative to the Cartesian notation expresses the
electrostatic potential using the notation of Morse and
Feshbach,\cite{Morse:1946zr}
\begin{equation} \label{eq:quad_phi} 
\Phi(x,y,z) = -\left[ a_{20} \frac{2 z^2 -x^2 - y^2}{2}
+ 3 a_{21}^e \,xz + 3 a_{21}^o \,yz  
 + 6a_{22}^e \,xy +  3 a_{22}^o (x^2 - y^2) \right].
\end{equation}
Here we use the standard $(l,m)$ form for the $a_{lm}$ coefficients,
with superscript $e$ and $o$ denoting even and odd, respectively.
This form makes the functional analogy to ``d'' atomic states
apparent. 

Applying the gradient operator to Eq. (\ref{eq:quad_phi}) the electric
field due to this potential,
\begin{equation}
\mathbf{E} = -\nabla \Phi
= \left(\begin{array}{ccc}
\left( 6a_{22}^o -a_{20} \right)\; x &+\; 6a_{22}^e\; y &+\; 3a_{21}^e\;  z  \\
6a_{22}^e\; x & -\; (a_{20} + 6a_{22}^o)\; y & +\; 3a_{21}^o\; z \\
3a_{21}^e\; x & +\; 3a_{21}^o\; y & +\; 2a_{20}\; z
\end{array} \right),
\label{eq:MFE}
\end{equation}
while the gradient of the electric field in this form is:
\begin{equation} \label{eq:grad_e2}
\mathsf{G} = 
\begin{pmatrix}
6 a_{22}^o - a_{20} & 6a_{22}^e & 3a_{21}^e\\
6a_{22}^e & -(a_{20}+6a_{22}^o) & 3a_{21}^o \\
3a_{21}^e  &  3a_{21}^o & 2a_{20} \\
\end{pmatrix} \\
\end{equation}
which is also uniform over the entire space.  This form for the
gradient can be factored as
\begin{gather}
\begin{aligned}
\mathsf{G}  = a_{20} 
\begin{pmatrix}
-1 & 0 & 0\\
0 & -1 & 0\\
0 & 0 & 2\\
\end{pmatrix}
+3a_{21}^e
\begin{pmatrix}
0 & 0 & 1\\
0 & 0 & 0\\
1 & 0 & 0\\
\end{pmatrix}
+3a_{21}^o
\begin{pmatrix}
0 & 0 & 0\\
0 & 0 & 1\\
0 & 1 & 0\\
\end{pmatrix}
+6a_{22}^e
\begin{pmatrix}
0 & 1 & 0\\
1 & 0 & 0\\
0 & 0 & 0\\
\end{pmatrix}
+6a_{22}^o
\begin{pmatrix}
1 & 0 & 0\\
0 & -1 & 0\\
0 & 0 & 0\\
\end{pmatrix}.
\end{aligned}
 \label{eq:intro_tensors}
\end{gather}
The five matrices in the expression above represent five different
symmetric traceless tensors of rank 2. 

It is useful to find the Cartesian vectors $\mathbf a$ and $\mathbf b$
that generate the five types of tensors shown in
Eq. (\ref{eq:intro_tensors}).  If the two vectors are co-linear, e.g.,
$\psi=0$, $\mathbf{a}=(0,0,1)$ and $\mathbf{b}=(0,0,1)$, then
\begin{equation*}
\mathsf{G} = \frac{g_0}{3}
\begin{pmatrix}
-1 & 0 & 0 \\
0 & -1 & 0 \\
0 & 0 & 2 \\
\end{pmatrix} ,
\end{equation*}
which is the $a_{20}$ symmetry.
To generate the $a_{22}^o$ symmetry, we take:
$\mathbf{a}= (\frac{1}{\sqrt{2}}, \frac{1}{\sqrt{2}},0)$ and
$\mathbf{b}=(\frac{1}{\sqrt{2}}, -\frac{1}{\sqrt{2}},0)$
and find:
\begin{equation*}
\mathsf{G}=\frac{g_0}{2}
\begin{pmatrix}
1 & 0 & 0 \\
0 & -1 & 0 \\
0 & 0 & 0 \\
\end{pmatrix} .
\end{equation*}
To generate the $a_{22}^e$ symmetry, we take:
$\mathbf{a}= (1, 0, 0)$ and $\mathbf{b} = (0,1,0)$ and find:
\begin{equation*}
\mathsf{G}=\frac{g_0}{2}
\begin{pmatrix}
0 & 1 & 0 \\
1 & 0 & 0 \\
0 & 0 & 0 \\
\end{pmatrix} .
\end{equation*}
The pattern is straightforward to continue for the other symmetries.

We find the notation of Ref. \onlinecite{Torres-del-Castillo:2006uo}
helpful when creating specific types of constant gradient electric
fields in simulations. For this reason,
Eqs. (\ref{eq:appliedPotential}), (\ref{eq:CE}), and (\ref{eq:GC}) are
implemented in our code.  In the simulations using constant applied
gradients that are mentioned in the main text, we utilized a field
with the $a_{22}^e$ symmetry using vectors, $\mathbf{a}= (1, 0, 0)$
and $\mathbf{b} = (0,1,0)$.

\section{Point-multipolar interactions with a spatially-varying electric field}

This section develops formulas for the force and torque exerted by an
external electric field, $\mathbf{E}(\mathbf{r})$, on object
$a$.\cite{Raab:2004ve} Object $a$ has an embedded collection of
charges and in simulations will represent a molecule, ion, or a
coarse-grained substructure. We describe the charge distributions
using primitive multipoles defined in Ref. \onlinecite{PaperI} by
\begin{align}
C_a =&\sum_{k \, \text{in }a} q_k , \label{eq:charge} \\
D_{a\alpha} =&\sum_{k \, \text{in }a} q_k r_{k\alpha}, \label{eq:dipole}\\
Q_{a\alpha\beta} =& \frac{1}{2} \sum_{k \, \text{in }  a} q_k
r_{k\alpha}  r_{k\beta},
\label{eq:quadrupole}
\end{align}
where $\mathbf{r}_k$ is the local coordinate system for the object
(usually the center of mass of object $a$).  Components of vectors and
tensors are given using the Einstein repeated summation notation. Note
that the definition of the primitive quadrupole here differs from the
standard traceless form, and contains an additional Taylor-series
based factor of $1/2$. In Ref.  \onlinecite{PaperI}, we derived the
forces and torques each object exerts on the other objects in the
system.

Here we must also consider an external electric field that varies in
space: $\mathbf E(\mathbf r)$.  Each of the local charges $q_k$ in
object $a$ will then experience a slightly different field.  This
electric field can be expanded in a Taylor series around the local
origin of each object. For a particular charge $q_k$, the electric
field at that site's position is given by:
\begin{equation}
\mathbf{E}(\mathbf{r}_k) = E_\gamma|_{\mathbf{r}_k = 0} + \nabla_\delta E_\gamma |_{\mathbf{r}_k = 0}  r_{k \delta} 
+ \frac {1}{2} \nabla_\delta \nabla_\varepsilon E_\gamma|_{\mathbf{r}_k = 0}  r_{k \delta}
r_{k \varepsilon} + ... 
\end{equation}
Note that if one shrinks object $a$ to a single point, the
${E}_\gamma$ terms are all evaluated at the center of the object (now
a point). Thus later the ${E}_\gamma$ terms can be written using the
same (molecular) origin for all point charges in the object. The force
exerted on object $a$ by the electric field is given by,
\begin{align}
F^a_\gamma = \sum_{k \textrm{~in~} a} q_k E_\gamma(\mathbf{r}_k) &=  \sum_{k \textrm{~in~} a} q_k \lbrace E_\gamma + \nabla_\delta E_\gamma r_{k \delta} 
+ \frac {1}{2} \nabla_\delta \nabla_\varepsilon E_\gamma r_{k \delta}
r_{k \varepsilon} + ...  \rbrace  \\
 &= C_a E_\gamma + D_{a  \delta} \nabla_\delta E_\gamma 
+ Q_{a \delta \varepsilon} \nabla_\delta \nabla_\varepsilon E_\gamma +
... 
\end{align}
Thus in terms of the global origin $\mathbf{r}$, ${F}_\gamma(\mathbf{r}) = C {E}_\gamma(\mathbf{r})$ etc. 
  
Similarly, the torque exerted by the field on $a$ can be expressed as
\begin{align}
\tau^a_\alpha &=  \sum_{k \textrm{~in~} a} (\mathbf r_k \times q_k \mathbf E)_\alpha \\
 & =  \sum_{k \textrm{~in~} a} \epsilon_{\alpha \beta \gamma} q_k
 r_{k\beta} E_\gamma(\mathbf r_k) \\
 & = \epsilon_{\alpha \beta \gamma} D_\beta E_\gamma 
+ 2 \epsilon_{\alpha \beta \gamma} Q_{\beta \delta} \nabla_\delta
E_\gamma + ...
\end{align}
We note that the Levi-Civita symbol can be eliminated by utilizing the matrix cross product as defined in Ref. \onlinecite{Smith98}:
\begin{equation}
\left[\mathsf{A} \times \mathsf{B}\right]_\alpha = \sum_\beta
\left[\mathsf{A}_{\alpha+1,\beta} \mathsf{B}_{\alpha+2,\beta}
  -\mathsf{A}_{\alpha+2,\beta} \mathsf{B}_{\alpha+1,\beta} 
\right]
\label{eq:matrixCross}
\end{equation}
where $\alpha+1$ and $\alpha+2$ are regarded as cyclic permuations of
the matrix indices. Finally, the interaction energy $U^a$ of object $a$ with the external field is given by,
\begin{equation}
U^a = \sum_{k~in~a} q_k \phi_k (\mathrm{r}_k)
\end{equation}
Performing another Taylor series expansion about the local body origin,
\begin{equation}
\phi({\mathbf{r}_k}) = \phi|_{\mathbf{r}_k = 0 } + r_{k \alpha} \nabla_\alpha \phi_\alpha|_{\mathbf{r}_k = 0 } + \frac{1}{2} r_{k\alpha}r_{k\beta}\nabla_\alpha \nabla_\beta \phi|_{\mathbf{r}_k = 0} + ...
\end{equation}
Writing this in terms of the global origin $\mathbf{r}$, we find
\begin{equation}
U(\mathbf{r}) = \mathrm{C} \phi(\mathbf{r}) - \mathrm{D}_\alpha \mathrm{E}_\alpha - \mathrm{Q}_{\alpha\beta}\nabla_\alpha \mathrm{E}_\beta + ...
\end{equation}
These results have been summarized in Table \ref{tab:UFT}.

\begin{table}
\caption{Potential energy $(U)$, force $(\mathbf{F})$, and torque
  $(\mathbf{\tau})$ expressions for a multipolar site at $\mathbf{r}$ in an
  electric field, $\mathbf{E}(\mathbf{r})$ using the definitions of the multipoles in Eqs. (\ref{eq:charge}), (\ref{eq:dipole}) and (\ref{eq:quadrupole}).  
  \label{tab:UFT}}
\begin{tabular}{r|C{3cm}C{3cm}C{3cm}}
  & Charge & Dipole & Quadrupole \\ \hline
$U(\mathbf{r})$ &  $C \phi(\mathbf{r})$ & $-\mathbf{D} \cdot \mathbf{E}(\mathbf{r})$ & $- \mathsf{Q}:\nabla \mathbf{E}(\mathbf{r})$ \\
$\mathbf{F}(\mathbf{r})$ & $C \mathbf{E}(\mathbf{r})$ & $\mathbf{D} \cdot \nabla \mathbf{E}(\mathbf{r})$ &  $\mathsf{Q} : \nabla\nabla\mathbf{E}(\mathbf{r})$ \\
$\mathbf{\tau}(\mathbf{r})$ & & $\mathbf{D} \times \mathbf{E}(\mathbf{r})$ & $2 \mathsf{Q} \times \nabla \mathbf{E}(\mathbf{r})$
\end{tabular}
\end{table}

\section{Boltzmann averages for orientational polarization}
If we consider a collection of molecules in the presence of external
field, the perturbation experienced by any one molecule will include
contributions to the field or field gradient produced by the all other
molecules in the system. In subsections
\ref{subsec:boltzAverage-Dipole} and \ref{subsec:boltzAverage-Quad},
we discuss the molecular polarization due solely to external field
perturbations.  This illustrates the origins of the polarizability
equations (Eqs. 6, 20, and 21) in the main text.

\subsection{Dipoles}
\label{subsec:boltzAverage-Dipole}
Consider a system of molecules, each with permanent dipole moment
$p_o$. In the absense of an external field, thermal agitation orients
the dipoles randomly, and the system moment, $\mathbf{P}$, is zero.
External fields will line up the dipoles in the direction of applied
field.  Here we consider the net field from all other molecules to be
zero.  Therefore the total Hamiltonian acting on each molecule
is,\cite{Jackson98}
\begin{equation}
H = H_o - \mathbf{p}_o \cdot \mathbf{E},
\end{equation}
where $H_o$ is a function of the internal coordinates of the molecule.
The Boltzmann average of the dipole moment in the direction of the
field is given by,
\begin{equation}
\langle p_{mol} \rangle = \frac{\displaystyle\int p_o \cos\theta
  e^{~p_o E \cos\theta /k_B T}\; d\Omega}{\displaystyle\int  e^{~p_o E \cos\theta/k_B
    T}\; d\Omega},
\end{equation}
where the $z$-axis is taken in the direction of the applied field,
$\bf{E}$ and
$\int d\Omega = \int_0^\pi \sin\theta\; d\theta \int_0^{2\pi} d\phi
\int_0^{2\pi} d\psi$
is an integration over Euler angles describing the orientation of the
molecule.

If the external fields are small, \textit{i.e.}
$p_oE \cos\theta / k_B T << 1$,
\begin{equation}
\langle p_{mol} \rangle \approx \frac{{p_o}^2}{3 k_B T}E,
\end{equation}
where $ \alpha_p = \frac{{p_o}^2}{3 k_B T}$ is the molecular
polarizability. The orientational polarization depends inversely on
the temperature as the applied field must overcome thermal agitation
to orient the dipoles.

\subsection{Quadrupoles}
\label{subsec:boltzAverage-Quad}
If instead, our system consists of molecules with permanent
\textit{quadrupole} tensor $q_{\alpha\beta}$. The average quadrupole
at temperature $T$ in the presence of uniform applied field gradient
is given by,\cite{AduGyamfi78, AduGyamfi81}
\begin{equation}
\langle q_{\alpha\beta} \rangle \;=\; \frac{\displaystyle\int
  q_{\alpha\beta}\; e^{-H/k_B T}\; d\Omega}{\displaystyle\int
  e^{-H/k_B T}\; d\Omega} \;=\; \frac{\displaystyle\int
  q_{\alpha\beta}\; e^{~q_{\mu\nu}\;\partial_\nu E_\mu /k_B T}\;
  d\Omega}{\displaystyle\int  e^{~q_{\mu\nu}\;\partial_\nu E_\mu /k_B
    T}\; d\Omega },
\label{boltzQuad}
\end{equation}
where $H = H_o - q_{\mu\nu}\;\partial_\nu E_\mu $ is the energy of a
quadrupole in the gradient of the applied field and $H_o$ is a
function of internal coordinates of the molecule. The energy and
quadrupole moment can be transformed into the body frame using a
rotation matrix $\mathsf{\eta}^{-1}$,
\begin{align}
q_{\alpha\beta} &= \eta_{\alpha\alpha'}\;\eta_{\beta\beta'}\;{q}^* _{\alpha'\beta'} \\
H &= H_o - q:{\nabla}\mathbf{E} \\
  &= H_o - q_{\mu\nu}\;\partial_\nu E_\mu  \\
  &= H_o
    -\eta_{\mu\mu'}\;\eta_{\nu\nu'}\;{q}^*_{\mu'\nu'}\;\partial_\nu
    E_\mu. \label{energyQuad}
\end{align}
Here the starred tensors are the components in the body fixed
frame. Substituting equation (\ref{energyQuad}) in the equation
(\ref{boltzQuad}) and taking linear terms in the expansion we obtain,
\begin{equation}
\braket{q_{\alpha\beta}} = \frac{\displaystyle \int q_{\alpha\beta} \left(1 +
    \frac{\eta_{\mu\mu'}\;\eta_{\nu\nu'}\;{q}^*_{\mu'\nu'}\;\partial_\nu
      E_\mu }{k_B T}\right)\;  d\Omega}{\displaystyle \int \left(1 + \frac{\eta_{\mu\mu'}\;\eta_{\nu\nu'}\;{q}^*_{\mu'\nu'}\;\partial_\nu E_\mu }{k_B T}\right)\; d\Omega}.
\end{equation}
Recall that $\eta_{\alpha\alpha'}$ is the inverse of the rotation
matrix that transforms the body fixed coordinates to the space
coordinates.
% \[\eta_{\alpha\alpha'} 
% = \left(\begin{array}{ccc}
% cos\phi\; cos\psi - cos\theta\; sin\phi\; sin\psi & -cos\theta\; cos\psi\; sin\phi - cos\phi\; sin\psi & sin\theta\; sin\phi \\
% cos\psi\; sin\phi + cos\theta\; cos\phi \; sin\psi & cos\theta\; cos\phi\; cos\psi - sin\phi\; sin\psi & -cos\phi\; sin\theta \\
% sin\theta\; sin\psi & -cos\psi\; sin\theta & cos\theta
% \end{array} \right).\]

Integration of the first and second terms in the denominator gives
$8 \pi^2$ and
$8 \pi^2 ({\nabla} \cdot \mathbf{E}) \mathrm{Tr}(q^*) / 3 $
respectively. The second term vanishes for charge free space (where
${\nabla} \cdot \mathbf{E}=0$). Similarly, integration of the first
term in the numerator produces
$8 \pi^2 \delta_{\alpha\beta} \mathrm{Tr}(q^*) / 3$ while the second
produces
$8 \pi^2 (3{q}^*_{\alpha'\beta'}{q}^*_{\beta'\alpha'} -
{q}^*_{\alpha'\alpha'}{q}^*_{\beta'\beta'})\partial_\alpha E_\beta /
15 k_B T $.
Therefore the Boltzmann average of a quadrupole moment can be written
as,
\begin{equation}
\langle q_{\alpha\beta} \rangle =  \frac{1}{3} \mathrm{Tr}(q^*)\;\delta_{\alpha\beta} + \frac{{\bar{q_o}}^2}{15k_BT}\;\partial_\alpha E_\beta,
\end{equation}
where $\alpha_q = \frac{{\bar{q_o}}^2}{15k_BT} $ is a molecular
quadrupole polarizablity and
${\bar{q_o}}^2=
3{q}^*_{\alpha'\beta'}{q}^*_{\beta'\alpha'}-{q}^*_{\alpha'\alpha'}{q}^*_{\beta'\beta'}$
is the square of the net quadrupole moment of a molecule.

\section{Gradient of the field due to quadrupolar polarization}
\label{singularQuad}
In section IV.C of the main text, we stated that for quadrupolar
fluids, the self-contribution to the field gradient vanishes at the
singularity. In this section, we prove this statement.  For this
purpose, we consider a distribution of charge $\rho(\mathbf{r})$ which
gives rise to an electric field $\mathbf{E}(\mathbf{r})$ and gradient
of the field $\nabla\mathbf{E}(\mathbf{r})$ throughout space. The
gradient of the electric field over volume due to the charges within
the sphere of radius $R$ is given by (cf. Ref. \onlinecite{Jackson98},
equation 4.14):
\begin{equation}
\int_{r<R} \nabla\mathbf{E} d\mathbf{r} = -\int_{r=R} R^2 \mathbf{E}\;\hat{n}\; d\Omega
\label{eq:8}
\end{equation}
where $d\Omega$ is the solid angle and $\hat{n}$ is the normal vector
of the surface of the sphere, 
\begin{equation}
\hat{n} = \sin\theta\cos\phi\; \hat{x} + \sin\theta\sin\phi\; \hat{y} +
\cos\theta\; \hat{z}
\end{equation}
in spherical coordinates.  For the charge density $\rho(\mathbf{r}')$, the
total gradient of the electric field can be written as,\cite{Jackson98}
\begin{equation}
\int_{r<R} {\nabla}\mathbf {E}\; d\mathbf{r}=-\int_{r=R} R^2\;
{\nabla}\Phi\; \hat{n}\; d\Omega
=-\frac{1}{4\pi\;\epsilon_o}\int_{r=R} R^2\; {\nabla}\;\left(\int
  \frac{\rho(\mathbf
    r')}{|\mathbf{r}-\mathbf{r}'|}\;d\mathbf{r}'\right) \hat{n}\;
d\Omega .
\label{eq:9}
\end{equation}
The radial function in the equation (\ref{eq:9}) can be expressed in
terms of spherical harmonics as,\cite{Jackson98}
\begin{equation}
\frac{1}{|\mathbf{r} - \mathbf{r}'|} = 4\pi \sum_{l=0}^{\infty}\sum_{m=-l}^{m=l}\frac{1}{2l+1}\;\frac{{r^l_<}}{{r^{l+1}_>}}\;{Y^*}_{lm}(\theta', \phi')\;Y_{lm}(\theta, \phi)
\label{eq:10}
\end{equation}
If the sphere completely encloses the charge density then $ r_< = r'$ and $r_> = R$. Substituting equation (\ref{eq:10}) into (\ref{eq:9}) we get,
\begin{equation}
\begin{split}
\int_{r<R} {\nabla}\mathbf{E}\;d\mathbf{r} &=-\frac{R^2}{\epsilon_o}\int_{r=R} \; {\nabla}\;\left(\int \rho(\mathbf r')\sum_{l=0}^{\infty}\sum_{m=-l}^{m=l}\frac{1}{2l+1}\;\frac{{r'^l}}{{R^{l+1}}}\;{Y^*}_{lm}(\theta', \phi')\;Y_{lm}(\theta, \phi)\;d\mathbf{r}'\right) \hat{n}\; d\Omega \\
 &= -\frac{R^2}{\epsilon_o}\sum_{l=0}^{\infty}\sum_{m=-l}^{m=l}\frac{1}{2l+1}\;\int \rho(\mathbf r')\;{r'^l}\;{Y^*}_{lm}(\theta', \phi')\left(\int_{r=R}\vec{\nabla}\left({R^{-(l+1)}}\;Y_{lm}(\theta, \phi)\right)\hat{n}\; d\Omega \right)d\mathbf{r}
' .
\end{split}
\label{eq:11}
\end{equation} 
The gradient of the product of radial function and spherical harmonics
is given by:\cite{Arfkan}
\begin{equation}
\begin{split}
{\nabla}\left[ f(r)\;Y_{lm}(\theta, \phi)\right] = &-\left(\frac{l+1}{2l+1}\right)^{1/2}\; \left[\frac{\partial}{\partial r}-\frac{l}{r} \right]f(r)\; Y_{l, l+1, m}(\theta, \phi)\\ &+ \left(\frac{l}{2l+1}\right)^{1/2}\left[\frac
{\partial}{\partial r}+\frac{l}{r} \right]f(r)\; Y_{l, l-1, m}(\theta, \phi).
\end{split}
\label{eq:12}
\end{equation}
where $Y_{l,l+1,m}(\theta, \phi)$ is a vector spherical
harmonic.\cite{Arfkan} Using equation (\ref{eq:12}) we get,
\begin{equation}
{\nabla}\left({R^{-(l+1)}}\;Y_{lm}(\theta, \phi)\right) = [(l+1)(2l+1)]^{1/2}\; Y_{l,l+1,m}(\theta, \phi) \; \frac{1}{R^{l+2}},
\label{eq:13}
\end{equation}
Using Clebsch-Gordan coefficients $C(l+1,1,l|m_1,m_2,m)$, the vector
spherical harmonics can be written in terms of spherical harmonics,
\begin{equation}
Y_{l,l+1,m}(\theta, \phi) = \sum_{m_1, m_2} C(l+1,1,l|m_1,m_2,m)\; Y_{l+1}^{m_1}(\theta,\phi)\; \hat{e}_{m_2}.
\label{eq:14}
\end{equation}
Here $\hat{e}_{m_2}$ is a spherical tensor of rank 1 which can be expressed
in terms of Cartesian coordinates,
\begin{equation}
{\hat{e}}_{+1} = - \frac{\hat{x}+i\hat{y}}{\sqrt{2}},\quad {\hat{e}}_{0} = \hat{z},\quad and \quad {\hat{e}}_{-1} = \frac{\hat{x}-i\hat{y}}{\sqrt{2}}.
\label{eq:15}
\end{equation} 
The normal vector $\hat{n} $ is then expressed in terms of spherical tensor of rank 1 as shown in below,
\begin{equation}
\hat{n} = \sqrt{\frac{4\pi}{3}}\left(-Y_1^{-1}{\hat{e}}_1 - Y_1^{1}{\hat{e}}_{-1} + Y_1^{0}{\hat{e}}_0 \right).
\label{eq:16}
\end{equation}
The surface integral of the product of $\hat{n}$ and
$Y_{l+1}^{m_1}(\theta, \phi)$ gives,
\begin{equation}
\begin{split}
\int \hat{n}\;Y_{l+1}^{m_1}\;d\Omega &= \int \sqrt{\frac{4\pi}{3}}\left(-Y_1^{-1}{\hat{e}}_1 -Y_1^{1}{\hat{e}}_{-1} + Y_1^{0}{\hat{e}}_0 \right)\;Y_{l+1}^{m_1}\; d\Omega \\
&=  \int \sqrt{\frac{4\pi}{3}}\left({Y_1^{1}}^* {\hat{e}}_1 +{Y_1^{-1}}^* {\hat{e}}_{-1} + {Y_1^{0}}^* {\hat{e}}_0 \right)\;Y_{l+1}^{m_1}\; d\Omega \\
&=   \sqrt{\frac{4\pi}{3}}\left({\delta}_{l+1, 1}\;{\delta}_{1, m_1}\;{\hat{e}}_1 + {\delta}_{l+1, 1}\;{\delta}_{-1, m_1}\;{\hat{e}}_{-1}+ {\delta}_{l+1, 1}\;{\delta}_{0, m_1} \;{\hat{e}}_0\right),
\end{split}
\label{eq:17}
\end{equation}
where $Y_{l}^{-m} = (-1)^m\;{Y_{l}^{m}}^* $ and
$ \int {Y_{l}^{m}}^* Y_{l'}^{m'}\;d\Omega =
\delta_{ll'}\delta_{mm'} $.
Non-vanishing values of equation \ref{eq:17} require $l = 0$,
therefore the value of $ m = 0 $. Since the values of $ m_1$ are -1,
1, and 0 then $m_2$ takes the values 1, -1, and 0, respectively
provided that $m = m_1 + m_2$.  Equation \ref{eq:11} can therefore be
modified,
\begin{equation}
\begin{split}
\int_{r<R} {\nabla}\mathbf{E}\;d\mathbf{r} = &- \sqrt{\frac{4\pi}{{3}}}\;\frac{1}{\epsilon_o}\int \rho(r')\;{Y^*}_{00}(\theta', \phi')[ C(1, 1, 0|-1,1,0)\;{\hat{e}_{-1}}{\hat{e}_{1}}\\  &+ C(1, 1, 0|-1,1,0)\;{\hat{e}_{1}}{\hat{e}_{-1}}+C(
1, 1, 0|0,0,0)\;{\hat{e}_{0}}{\hat{e}_{0}} ]\; d\mathbf{r}' \\
&= -\sqrt{\frac{4\pi}{{3}}}\;\frac{1}{\epsilon_o}\int \rho(r')\;d\mathbf{r}'\left({\hat{e}_{-1}}{\hat{e}_{1}}+{\hat{e}_{1}}{\hat{e}_{-1}}-{\hat{e}_{0}}{\hat{e}_{0}}\right)\\
&= - \sqrt{\frac{4\pi}{{3}}}\;\frac{1}{\epsilon_o}\;C_\mathrm{total}\;\left({\hat{e}_{-1}}{\hat{e}_{1}}+{\hat{e}_{1}}{\hat{e}_{-1}}-{\hat{e}_{0}}{\hat{e}_{0}}\right).
\end{split}
\label{eq:19} 
\end{equation}
In the last step, the charge density was integrated over the sphere,
yielding a total charge $C_\mathrm{total}$. Equation (\ref{eq:19})
gives the total gradient of the field over a sphere due to the
distribution of the charges.  For quadrupolar fluids the total charge
within a sphere is zero, therefore
$ \int_{r<R} {\nabla}\mathbf{E}\;d\mathbf{r} = 0 $.  Hence the quadrupolar
polarization produces zero net gradient of the field inside the
sphere.

\section{Corrections to the distance-dependent Kirkwood function}
\label{kirkwood}
\begin{figure}
\includegraphics[width=\linewidth]{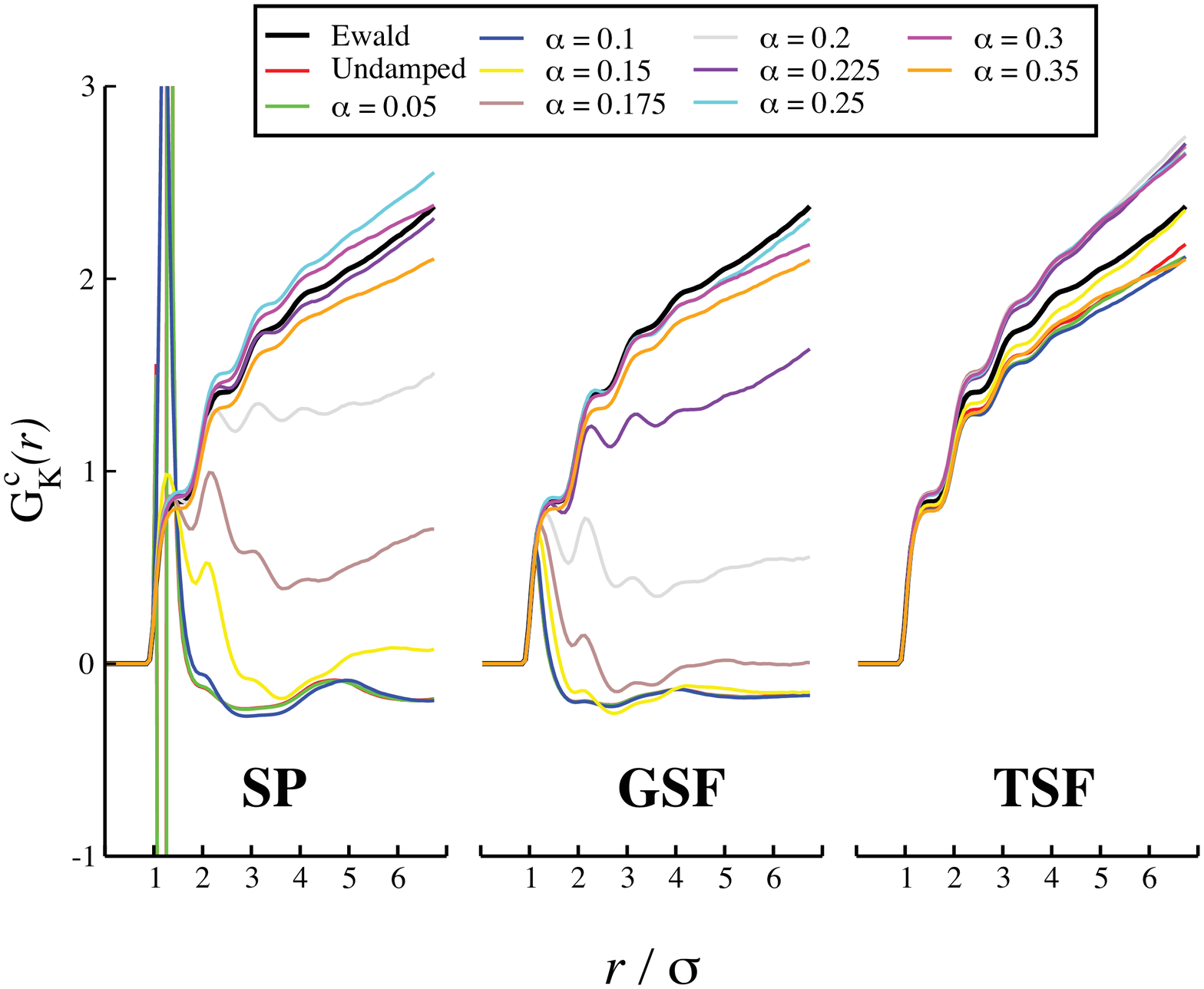}
\caption{Corrected distance-dependent factors of the dipolar system
  for the three real space methods at a range of Gaussian damping
  parameters ($\alpha$) with a cutoff
  $r_c = 3.52 \sigma = 12$~\AA$^{-1}$.}
\label{fig:kirkwoodCorr}
\end{figure}

In the main text, we provide data on the distance-dependent Kirkwood
function, 
\begin{equation}
G_K(r) = \left< \frac{1}{N} \sum_{i} \sum_{\substack{j \\ r_{ij} < r}}
  \frac{\mathbf{D}_i \cdot \mathbf{D}_j}{\left| D_i \right| \left| D_j
    \right|} \right>,
\label{eq:kirkwood}
\end{equation}
which is a sensitive measure of orientational ordering in dipolar
liquids.  We noted in the main text that because the Kirkwood function
is measuring the same bulk dipolar response as the dielectric
constant, it is possible to apply a correction for the truncation,
force shifting, and damping to arrive at a corrected Kirkwood
function. Starting with Eq. (16) in the main text and recognizing
$\epsilon = 1 + \chi_D$ and $\epsilon_{CB} = 1 + \alpha_D$, we arrive
at a relationship between the (corrected) susceptibility,
\begin{equation}
\chi_D = \frac{\alpha_D}{1 + (A-1) \frac{\alpha_D}{3}}.
\label{eq:chiAlpha}
\end{equation}
and the dipole polarizability, which is easily obtained from bulk
simulations, 
\begin{equation}
\alpha_D =\frac{\braket{\mathbf{M}^2}-{\braket{\mathbf{M}}}^2}{3
  \epsilon_o V k_B T}.
\end{equation}
In the absence of bulk dipolar ordering, ${\braket{\mathbf{M}}}^2=0$,
and we may recognize the polarizability as a limit of the
distance-dependent Kirkwood function,
\begin{equation}
\alpha_D = \left( \frac{\rho D^2}{3 \epsilon_0 k_B T}\right) \lim_{r
  \rightarrow \infty}
G_K(r).
\label{eq:alphaGK}
\end{equation}
Here we have used the number density, $\rho = N/V$ and the molecular
dipole moment, $D$, to connect with Eq. (\ref{eq:kirkwood}).  By
analogy, the dipolar susceptibility may be connected with a corrected
(but unknown) version of the Kirkwood function,
\begin{equation}
  \chi_D = \left(\frac{\rho D^2}{3 \epsilon_0 k_B T}\right) \lim_{r
    \rightarrow \infty}
  G_K^c(r).
\label{eq:chiGKc}
\end{equation}
Substituting the corrected and raw Kirkwood expressions,
Eqs. (\ref{eq:chiGKc}) and (\ref{eq:alphaGK}), into
Eq. (\ref{eq:chiAlpha}), and removing the limits, we obtain a
correction formula for the Kirkwood function,
\begin{equation}
G_K^c(r) = \frac{G_K(r)}{ 1 + (A-1) \frac{\rho D^2 G_K(r)}{9
    \epsilon_0 k_B T}}.
\label{eq:kirkwoodCorr}
\end{equation}
Note that this correction forumla uses the same $A$ parameter from
Table I in the main text.  As in case of the dielectric constant, the
Kirkwood correction is similarly sensitive to values of $A$ away from
unity.  In Fig.~\ref{fig:kirkwoodCorr} we show the corrected
$G_K^c(r)$ functions for the three real space methods with
$r_c = 3.52 \sigma = 12$~\AA~ and for the Ewald sum (with
$\kappa = 0.3119$~\AA$^{-1}$).

The correction does help reduce the effect of the ``hole'' in the
underdamped cases, particularly for the GSF method.  However, this
comes at the expense of a divergence when $G_K(r) \sim 1$ for the
underdamped SP case.  We find it more useful to look at the
uncorrected $G_K(r)$ functions to study orientational correlations,
which is why the uncorrected functions are shown in the main text.

\newpage
%\bibliography{dielectric_new}